\documentclass{aastex63}

\usepackage{graphicx}
\usepackage{amsmath}
\usepackage{enumitem}

\usepackage{epsfig}

\usepackage{subfigure}
\usepackage{longtable}
\usepackage{times}

\shorttitle{The gravitational collapse toward the Black Hole in
  Kerr and EGB Gravities}
\shortauthors{O.Donmez}

\begin{document}

\title{The gravitational collapse of the dust toward the newly formed rotating black holes in 
  Kerr and 4-D Einstein-Gauss-Bonnet Gravities}

\author{Orhan Donmez}
\altaffiliation{College of Engineering and Technology, American
  University of the Middle East, Kuwait}

\begin{abstract}
  Studying the gravitational collapse of dust particles toward newly formed black holes has
  gained popularity following the observation of gravitational waves resulting from the merger
  of black holes. In this paper, we focus on modeling the descent of dust debris toward a
  black hole using a numerical code that incorporates relativistic hydrodynamics in the
  framework of General and Einstein-Gauss Bonnet gravity. We explore the influence of
  various parameters, such as the black hole's rotation parameter $a$ and the EGB coupling constant
  $\alpha$, on the curvature effects observed. Both parameters significantly impact the
  dynamics of the accretion disk formed around the black holes. Furthermore, we discuss
  the gravitational collapsing process in two distinct scenarios. It is also observed that
  the mass accretion rate is significantly influenced by these two parameters. The rate at
  which mass is accreted toward a black hole directly impacts the black hole's growth and
  evolutionary trajectory.

\end{abstract}


%%%%%%%%%%%%%%%%%%%%%%%%%%%%%%%%%%%%%%%%%%%%%%%%%%%%%%%%%%%%%%%%%%%%%%%
%%%%%%%%%%%%%%%%%%%%%%%%%%%%%%%%%%%%%%%%%%%%%%%%%%%%%%%%%%%%%%%%%%%%%%%
%%%%%%%%%%%%%%%%%%%%%%%%%%%%%%%%%%%%%%%%%%%%%%%%%%%%%%%%%%%%%%%%%%%%%%%
%%%%%%%%%%%%%%%%%%%%%%%%%%%%%%%%%%%%%%%%%%%%%%%%%%%%%%%%%%%%%%%%%%%%%%%
%%%%%%%%%%%%%%%%%%%%%%%%%%%%%%%%%%%%%%%%%%%%%%%%%%%%%%%%%%%%%%%%%%%%%%%
%%%%%%%%%%%%%%%%%%%%%%%%%%%%%%%%%%%%%%%%%%%%%%%%%%%%%%%%%%%%%%%%%%%%%%%

\section{Introduction}
\label{Introduction}

It is well-established that supermassive black holes reside at the centers of most
galaxies in the Universe. These extremely dense objects are thought to have
accumulated their massive sizes over cosmic timescales through the process of
accretion, where they gather and consume surrounding material \citep{Merloni2013}.
The origin and formation process of supermassive black holes (SMBHs) located at the center of
galaxies are still not fully understood. There are important unanswered questions surrounding
their growth and physical formation. For instance, it remains unclear how they can reach such
large sizes and accumulate mass at such rapid rates, as observed in the most distant quasars
\citep{Woods1}. Additionally, the nature and mass of the black hole seeds
that eventually develop into SMBHs with masses around $10^8-10^9$ solar masses in the early
universe\citep{Zhu1} are not well known. Furthermore, the relationship between the total
mass of a host galaxy and the mass of its central SMBH is still a topic of investigation
\citep{Volonteri1}.

In the literature, different fueling mechanisms for supermassive black holes
can be broadly categorized into two groups: internal and external processes.
Internal fueling mechanisms are associated with the secular evolution of galaxies.
These mechanisms include disk instabilities, such as those described
by \citet{Hopkins2006} and \citet{Gatti2016},  stellar winds
\citep{Ciotti2007,Ciotti2012, Kauffmann2009}, or the biased collapse of baryons in the
inner region of the halo, as discussed by \citet{Lapi2018}.
These processes can lead to the inward flow of gas towards the central
regions of the galaxy, ultimately fueling the central black hole.
  On the other hand,  external feeding is a mechanism that arises as a
  result of the interaction of a galaxy
  with stars or other galaxies in its vicinity.
These mechanisms involve galaxy interactions,
as studied by \citet{Gatti2016}, cold gas
inflows as described by \citet{Bournaud2012, DeGraf2017}.
These different fueling mechanisms represent various ways in which
supermassive black holes can acquire mass. They encompass both internal
processes that arise from the galaxy's own evolution and external processes
driven by interactions with the surrounding environment. Understanding the
relative contributions of these mechanisms is crucial for comprehending the
growth and evolution of supermassive black holes and their connection to the
larger-scale structure of galaxies and clusters.

Accretion of matter towards a black hole and the behavior of matter in a strong gravitational field
  can be understood by studying Kerr and Einstein-Gauss Bonnet (EGB) gravities. Accretion disks around
  the non-rotating EGB black holes have been explored theoretically and numerically
  \citep{Liu1, Haydari2, Donmez3}. Similarly, accretion disks around
  the rotating EGB black holes have been investigated
  in \citet{Fard1, Donmez_EGB_Rot}, and their references. On the other hand, the accretion disks
  around Schwarzschild and Kerr black holes have been studied for many years by a wide range of
  researchers. The process of black hole mass growth can be better understood by solving the
  equations of relativistic hydrodynamics. \citep{Abramowicz2013, Narayan2013}.
These equations take into account the effects of general
relativity and describe the behavior of matter and energy in the presence of strong gravitational
fields near the black hole horizon. By solving these equations, we can study the complex
dynamics of accretion flows, including the inflow and outflow of matter,
the formation of accretion disks, and the release of energy through various mechanisms
such as jets and radiation \citep{Narayan2008}.
This enables a more comprehensive understanding of how black
holes can grow in mass and the physical processes involved in their accretion.

  Different solutions of Einstein's equations can be an important tool to better understand astrophysical
  systems and explain the physical mechanisms leading to observations. Therefore, using the EGB gravity
  along with Kerr gravity here can help us gain a better understanding of physical events. It also
  provides an opportunity for comparison with Kerr. Thus, explaining spherically symmetric accretion
  using EGB gravity is important.
  However, there are those who believe that EGB gravity has not been produced with a correct approach.
  Therefore, there are studies in the literature suggesting that solutions obtained using EGB gravity
  may not be accurate \citep{Metin2020EPJC, Arrechea2020PhRvL, Bonifacio2020PhRvD} .
  Nevertheless, it has been demonstrated in the literature that these solutions remain valid in the
  4D flattened theory \citep{Banerjee2021ApJ}. The obtained black hole solutions \citep{Glavan2020PhRvL}
  remain valid in this plane \citep{Hennigar2020JHEP, Casalino2021PDU}. Therefore, the spherically
  symmetric solution obtained using the flattened theory reveals that it is the same as the Kerr solution.

Following the formation of a black hole resulting from a supernova explosion,
the material in the surrounding supernova remnant remains influenced by
the gravitational field of the black hole \citep{2012gcss.book}. Subsequently,
some of these materials
undergoes gravitational attraction and as it converges towards the black hole,
it organizes itself into an accretion disk, characterized by a swirling
configuration composed of gas and dust that envelops the black hole.
In this paper, we introduce a numerical model that examines the gravitational
collapse of the remnants toward a newly formed black hole.
The model utilizes a
general relativistic hydrodynamic code using two different gravity scenarios.
By conducting simulations, we track the system's evolution over time and make
predictions regarding important outcomes, such as the accretion rate toward the
black hole and the dynamic properties of the resulting accretion disk.
The efficiency of accretion directly impacts the growth rate of a black hole,
as it determines how effectively matter can transfer its energy and angular
momentum to the black hole. In efficient accretion processes, a significant
portion of the infalling matter's energy and angular momentum is successfully
captured by the black hole, leading to a more rapid growth of its mass.

The subsequent sections of the paper are structured as follows:
In Section \ref{GRHE1}, we delve into the formulation of general relativistic
hydrodynamics equations, specifically within the context of a rotating black hole.
We explore the intricacies of incorporating the effects of general relativity into the
hydrodynamic equations, considering the space-time curvature caused by the rotating black hole.
Furthermore, we introduce two different coordinate systems that are particularly
relevant for studying hydrodynamics near a rotating black hole. These coordinate
systems offer distinct advantages and insights into the dynamics of the fluid
surrounding the black hole.
In section \ref{GRHE2}, to ensure a comprehensive understanding, we also provide a detailed explanation
of the initial conditions, which describe the state of the fluid at the beginning
of the simulation, and boundary conditions, which define the behavior of the
fluid at boundaries of the computational domain. These conditions are essential
for accurately simulating and studying the hydrodynamical behavior near a rotating black hole.
The section  \ref{Results} then proceeds to discuss the obtained results in a
systematic manner. It presents the evolution of the fluid dynamics over time,
highlighting key features and phenomena observed during the simulation. The
analysis includes quantitative measurements, such as density profiles, velocity
distributions, and mass accretion rates, which are compared and contrasted under
different conditions and scenarios.
The occurrence of instability and quasi-periodic oscillations in the
disk-black hole system is discussed in section \ref{QPO}.
This section focuses on analyzing and describing the instabilities
observed in the system and the resulting quasi-periodic oscillatory behavior. 
The section \ref{Conclusion} provides a
comprehensive summary of all the
findings obtained throughout the study. This section serves as a culmination
of research and presents a condensed overview of the key results and their significance.
In the entire paper, unless explicitly mentioned otherwise, the convention of adopting
geometrized units where c and G are set to 1 is followed.

%%%%%%%%%%%%%%%%%%%%%%%%%%%%%%%%%%%%%%%%%%%%%%%%%%%%%%%%%%%%%%%%%%%%%%%
%%%%%%%%%%%%%%%%%%%%%%%%%%%%%%%%%%%%%%%%%%%%%%%%%%%%%%%%%%%%%%%%%%%%%%%

\section{Rotating Black Hole Metric and General Relativistic Equations}
\label{GRHE1}

In the gravitational collapse process, a fluid, such as gas or dust, is
gravitationally pulled towards a massive object, such as a black hole,
and accreted toward it. This process is important in understanding how matter
interacts with black holes and other compact objects in the universe.
  The investigation of  the gravitational collapse  process of a perfect fluid in the
presence of rotating black holes, specifically the Kerr and Einstein-Gauss-Bonnet
(EGB) black holes is studied by solving General Relativistic
Hydrodynamical (GRH) equations in the curved background.
The perfect fluid stress-energy-momentum tensor is

\begin{eqnarray}
 T^{ab} = \rho h u^{a}u^{b} + P g^{ab},
\label{GREq1}
\end{eqnarray}

\noindent $\rho$, $p$, $h$,  $u^{a}$, and $g^{ab}$ are  the rest-mass density, the fluid pressure,
the specific enthalpy,  the $4-$ velocity of the fluid, and
the metric of the curved space-time, respectively. The indexes $a$, $b$ and $c$ go from $0$ to $3$.
To compare the dynamic evolution of the accretion disk around the rotating black hole,
two distinct coordinate systems are employed. These are Kerr black hole in Boyer-Lindquist coordinate and
the rotating black hole metric in $4D$ EGB gravity.
  Kerr black hole in Boyer-Lindquist coordinate is
  \citep{Misner1973, Schutz2009, Donmez6}

\begin{eqnarray}
  ds^2 = -\left(1-\frac{2Mr}{\sum^2}\right)dt^2 - \frac{4Mra}{\sum^2}sin^2\theta dt d\phi
  + \frac{\sum^2}{\Delta_1}dr^2 + \sum^2 d\theta^2 + \frac{A}{\sum^2}sin^2\theta d\phi^2
\label{GREq2}
\end{eqnarray}

\noindent where $\Delta_1 = r^2 - 2Mr +a^2$, and
$A = (r^2 + a^2)^2 - a^2\Delta sin^2\theta$.
The lapse function and shift vector of the Kerr metric are
$\tilde{\alpha} = (\sum^2 \Delta_1/A)^{1/2} $ and $\beta^i = (0,0,-2Mar/A)$.

  \noindent The rotating black hole metric in $4D$ EGB gravity is
  \citep{Ghosh2020PDU, Wei2021EPJP, Donmez_EGB_Rot, Donmezetal2022}

\begin{eqnarray}
  ds^2 &=& -\frac{\Delta_2 - a^2sin^2\theta}{\Sigma}dt^2 + \frac{\Sigma}{\Delta_2}dr^2 -
  2asin^2\theta\left(1- \frac{\Delta_2 - a^2sin^2\theta}{\Sigma}\right)dtd\phi + 
  \Sigma d\theta^2 + \nonumber \\
  && sin^2\theta\left[\Sigma +  a^2sin^2\theta \left(2- \frac{\Delta_2 -  
   a^2sin^2\theta}{\Sigma} \right)  \right]d\phi^2,
\label{GREq3}
\end{eqnarray}

\noindent
where $\Sigma$ and $\Delta_2$ read as $\Sigma = r^2 + a^2cos^2\theta$ and
$\Delta_2 = r^2 + a^2 + \frac{r^4}{2\alpha}\left(1 - \sqrt{1 + \frac{8 \alpha M}{r^3}} \right)$.
$a$, $\alpha$, and $M$ are spin parameter, Gauss-Bonnet coupling constant, and
mass of the black hole, respectively. By solving equations $\Delta_1=0$  and $\Delta_2=0$,
the horizons of the black holes were determined. The lapse function $\tilde{\alpha}$
and the shift vectors of the EGB metric are
$\tilde{\alpha} = \sqrt{\frac{a^2(1-f(r))^2}{r^2+a^2(2-f(r))} + f(r)}$ and $\beta^i =
(0,\frac{a r^2}{2\pi \alpha}\left(1 -\sqrt{1 + \frac{8 \pi \alpha M}{r^3}}\right),0)$,
respectively.  $f(r) = 1 + \frac{r^2}{2\alpha}\left(1 -
\sqrt{1 + \frac{8 \alpha M}{r^3}} \right).$

\noindent To numerically solve the GRH equations, it is necessary to express them in a conserved form.
\citep{Donmez1}:

\begin{eqnarray}
  \frac{\partial U}{\partial t} + \frac{\partial F^r}{\partial r} + \frac{\partial F^{\phi}}{\partial \phi}
  = S.
\label{GREq4}
\end{eqnarray}

\noindent The vectors $U$, $F^r$, $F^{\phi}$, and $S$ represent the conserved variables,
fluxes along the $r$ and $\phi$ directions, and sources, respectively. These conserved
variables are expressed in terms of the primitive variables as seen below,

\begin{eqnarray}
  U =
  \begin{pmatrix}
    D \\
    S_r \\
    S_{\phi} \\
    \tau
  \end{pmatrix}
  =
  \begin{pmatrix}
    \sqrt{\gamma}W\rho \\
    \sqrt{\gamma}h\rho W^2 v_r\\
    \sqrt{\gamma}h\rho W^2 v_{\phi}\\
    \sqrt{\gamma}(h\rho W^2 - P - W \rho)
    \end{pmatrix}.
\label{GREq5}
\end{eqnarray}

\noindent
  In the provided equation, the quantities are defined as follows:
  $W = (1 - \gamma_{a,b}v^i v^j)^{-1/2}$ represents the Lorentz factor,
$h = 1 + \epsilon + P/\rho$ denotes the enthalpy, $\epsilon$ represents the internal energy,
and $v^i = u^i/W + \beta^i$ represents the three-velocity of the fluid. The pressure of the
fluid is determined using the ideal gas equation of state. The three-metric $\gamma_{i,j}$
and its determinant $\gamma$ are calculated using the four-metric of rotating black holes.
Latin indices $i$ and $j$ range from $1$ to $3$. The flux and source terms can be computed
for any metric using the following equations,

\begin{eqnarray}
  \vec{F}^i =
  \begin{pmatrix}
    \tilde{\alpha}\left(v^i - \frac{1}{\tilde{\alpha}\beta^i}\right)D \\
    \tilde{\alpha}\left(\left(v^i - \frac{1}{\tilde{\alpha}\beta^i}\right)S_j + \sqrt{\gamma}P\delta^i_j\right)\\
    \tilde{\alpha}\left(\left(v^i - \frac{1}{\tilde{\alpha}\beta^i}\right)\tau  + \sqrt{\gamma}P v^i\right)
    \end{pmatrix},
\label{GREq6}
\end{eqnarray}

\noindent and,

\begin{eqnarray}
  \vec{S} =
  \begin{pmatrix}
    0 \\
    \tilde{\alpha}\sqrt{\gamma}T^{ab}g_{bc}\Gamma^c_{aj} \\
    \tilde{\alpha}\sqrt{\gamma}\left(T^{a0}\partial_{a}\tilde{\alpha} - \tilde{\alpha}T^{ab}\Gamma^0_{ab}\right)
   \end{pmatrix},
\label{GREq7}
\end{eqnarray}

\noindent where $\Gamma^c_{ab}$ is the Christoffel symbol.

%%%%%%%%%%%%%%%%%%%%%%%%%%%%%%%%%%%%%%%%%%%%%%%%%%%%%%%%%%%%%%%%%%%%%%%
%%%%%%%%%%%%%%%%%%%%%%%%%%%%%%%%%%%%%%%%%%%%%%%%%%%%%%%%%%%%%%%%%%%%%%%

\section{Initial Conditions, Boundary Conditions, and Assumptions}
\label{GRHE2}

The accretion disks around black holes emit radiation at different ranges of frequencies depending on
their proximity to the black hole. $X-$rays typically occur in regions where the gravitational force
is significant, usually for $r < 100M$, i.e., the region up to the horizon of the black hole.
This region is generally limited to the area up to the horizon of the black hole. It is the region we
have used in our numerical simulations. The self-gravity resulting from the interaction of matter particles
is negligible compared to the gravitational force exerted by the black hole on matter. In other words,
in regions close to the black hole, the perfect fluid equation of state is used as the equation defining matter
pressure. This is a well-established and widely accepted approach in the literature \citep{Rezzolla2013}.

To investigate the gravitational collapse toward the rotating Gauss-Bonnet black hole
and compare it with the Kerr black hole, the General Relativistic Hydrodynamical
(GRH) equations are solved on the equatorial plane using the code described in
\citet{Donmez1, Donmez5, Donmez2}. The pressure of the accreted matter is
determined using the standard $\Gamma$
law equation of state for a perfect fluid, where $P = (\Gamma - 1)\rho\epsilon$
with $\Gamma = 4/3$. The initial density and pressure profiles are adjusted to
ensure that the speed of sound is equal to $C_{\infty} = 0.1$ in the outer boundary 
of the computational domain. After setting the density as a constant value
($\rho = 10^{-4}$), the pressure is computed based on the perfect fluid equation
of state. Subsequently, numerical simulations are performed on the equatorial
plane by injecting the gas from outer boundary with $V^r = -0.01$, $V^{\phi}=0$, and $\rho=1$.

  In this work, our aim is to elucidate the impact of Kerr and EGB gravities on the
  resulting stable disk around the slowly, moderately, and rapidly rotating black holes.  
  To achieve this goal, different values of the rotation parameter, $a/M=0.28$, $0.55$, and $0.9$, are
  employed to reveal the dynamic structure of the accretion disk around black holes in Kerr and
  EGB gravities. Various positive and negative EGB coupling constant ($\alpha$) values are
  used corresponding to each $a/M$, selected based on the criteria described in Fig.1 of
  \citet{Donmez_EGB_Rot}. As seen from \citet{Donmez_EGB_Rot}, there are a specific number of
  $\alpha$ corresponding to the same $a/M$, as chosen according to the criteria
  outlined in Fig.1 of \citet{Donmez_EGB_Rot}.
  In the positive direction, the value of $\alpha$ varies between $0$ and $1$, while in the
  negative direction, it takes values up to around $-6$ depending on the black hole's
  rotation parameter. For $a/M = 0.9$, positive $\alpha$  values are limited to a
  specific narrow range.

The computational domain is discretized using uniformly spaced zones in both the
radial and angular directions, with $N_r=1024$ zones in the radial direction and
$N_{\phi}=256$ zones in the angular direction. The inner and outer boundaries of the
computational domain are positioned at $r_{min}=2.3M$ and $r_{max}=100M$, respectively,
in the radial direction. In the angular direction, the boundaries are set as $\phi_{min}=0$
and $\phi_{max}=2\pi$.
The code execution time $(t_{max} = 30000M)$ significantly exceeded the time $(\sim 5000M)$
required for the model to reach the steady state.
It has been verified that the qualitative outcomes of the numerical
solutions, such as the presence of quasi-periodic oscillations (QPOs) and instabilities,
the location of shocks, and the behavior of the accretion rates, are not significantly
influenced by the grid resolution\citep{Donmez3, Donmez_EGB_Rot, Donmezetal2022}

Ensuring the accurate treatment of boundaries is crucial to avoid unphysical solutions
in numerical simulations. For the inner radial boundary, an outflow boundary condition
is implemented, allowing gas to fall into the black hole through a simple zeroth-order
extrapolation. In the outer boundary, gas is continuously injected
with initial velocities and density mention above to account for the inflow.
Periodic boundary conditions are employed along the $\phi$-direction.

  Black hole accretion disks form as a result of matter falling toward the black hole.
  During the formation of this disk, matter also falls into the black hole, crossing the
  innermost stable orbit. This, in turn, leads to an increase in the black hole's mass.
  However, the timescale for the increase in the black hole's mass is significantly larger
  than the timescale of physical events resulting from the interaction between the black
  hole and the disk, especially in the region close to the black hole within our computational
  domain. In other words, the timescale for accretion, and consequently, the change in the black
  hole's mass, is much larger than the timescale associated with the dynamic structure formation
  of the disk or radiation processes. Therefore, the effect of the change in the black
  hole's mass on the calculations can be considered negligible \citep{Alcubierre10, Jong2022JCAP}.

%%%%%%%%%%%%%%%%%%%%%%%%%%%%%%%%%%%%%%%%%%%%%%%%%%%%%%%%%%%%%%%%%%%%%%%
%%%%%%%%%%%%%%%%%%%%%%%%%%%%%%%%%%%%%%%%%%%%%%%%%%%%%%%%%%%%%%%%%%%%%%%
\section{The Numerical Simulation of the accretion toward 4D EGB Rotating and Kerr Black Holes}
\label{Results}

By solving the equations of general relativistic hydrodynamics on the equatorial
plane, employing the Kerr and EGB black hole space-time metrics along with the initial and
the boundary conditions given in section \ref{GRHE2},
we analyze the structure of a recently formed accretion disk and its instability
properties. The investigation encompasses the exploration of diverse spin parameter
values $(a)$ and Gauss-Bonnet coupling constant values $(\alpha)$. The objective is to
gain insights into the impact of these parameters on the characteristics and behavior
of the accretion disk, including the amount of mass accreted into the black hole.
Additionally, the study investigates the growth rate of the black hole mass and its
relation to the instability properties of the accretion disk.

%%%%%%%%%%%%%%%%%%%%%%%%%%%%%%%%%%%%%%%%%%%%%%%%%%%%%%%%%%%%%%%%%%%%%%

\begin{figure*}
  \vspace{1cm}
  \center
  \psfig{file=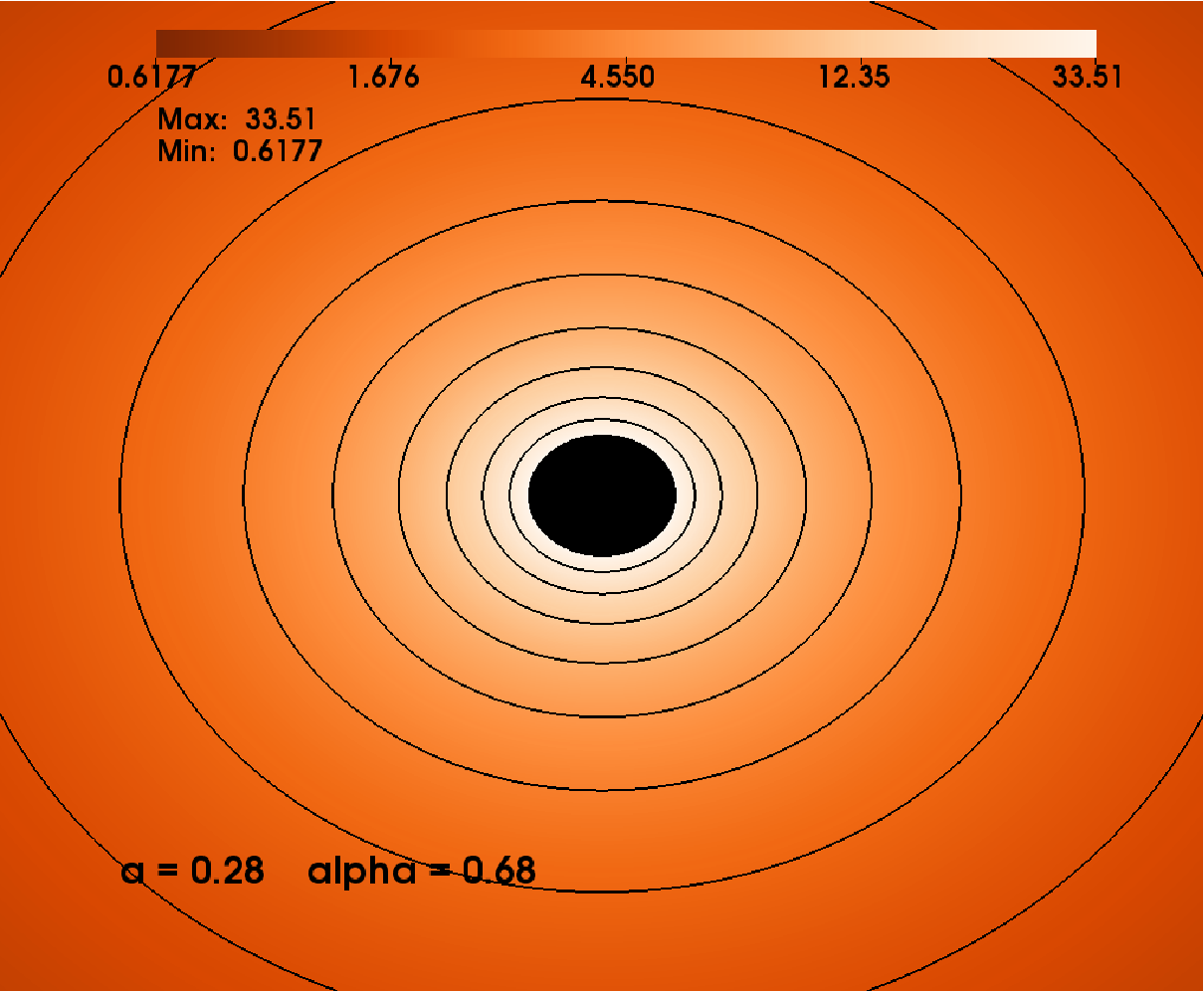,width=7.5cm}
  \psfig{file=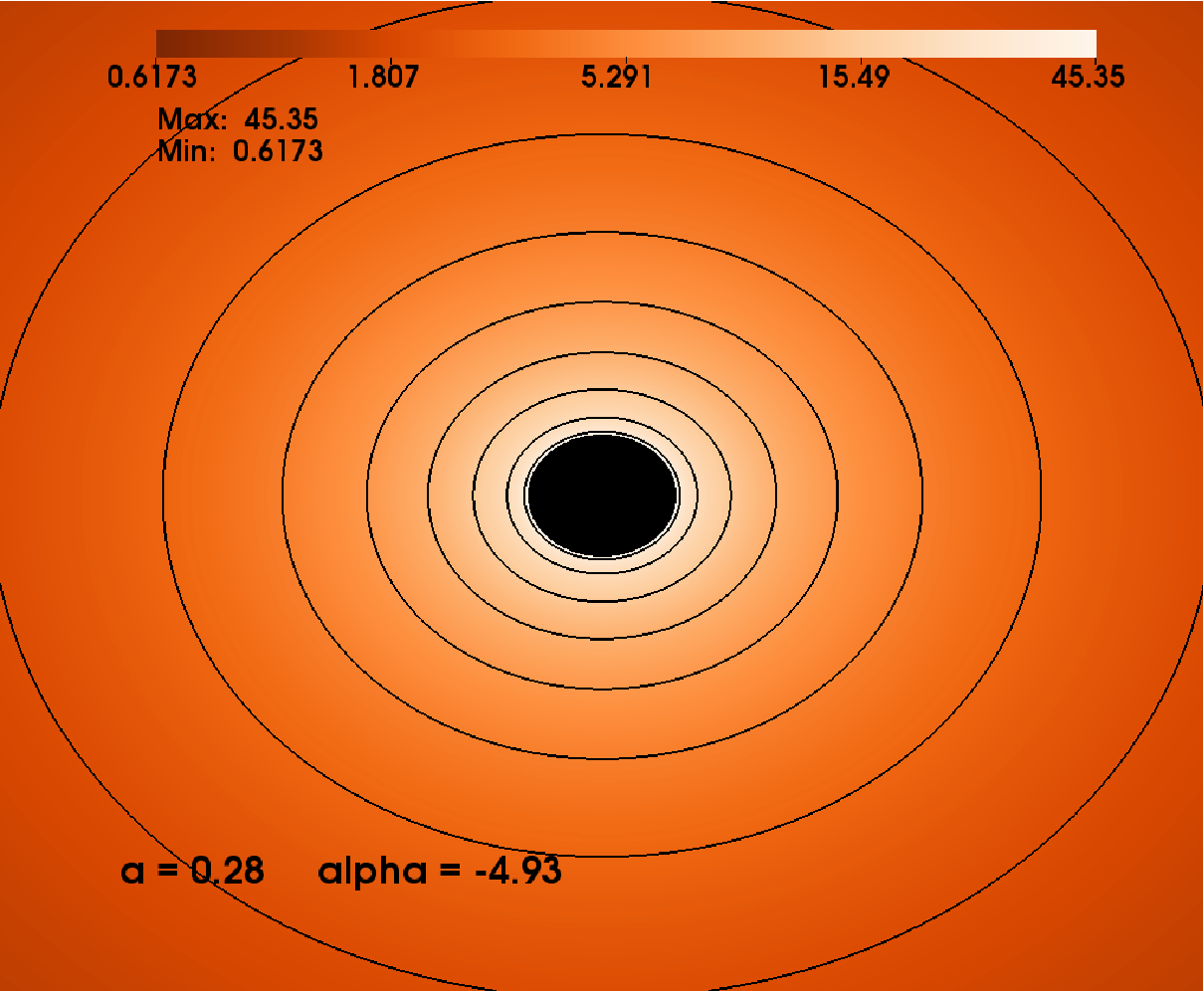,width=7.5cm}
  \psfig{file=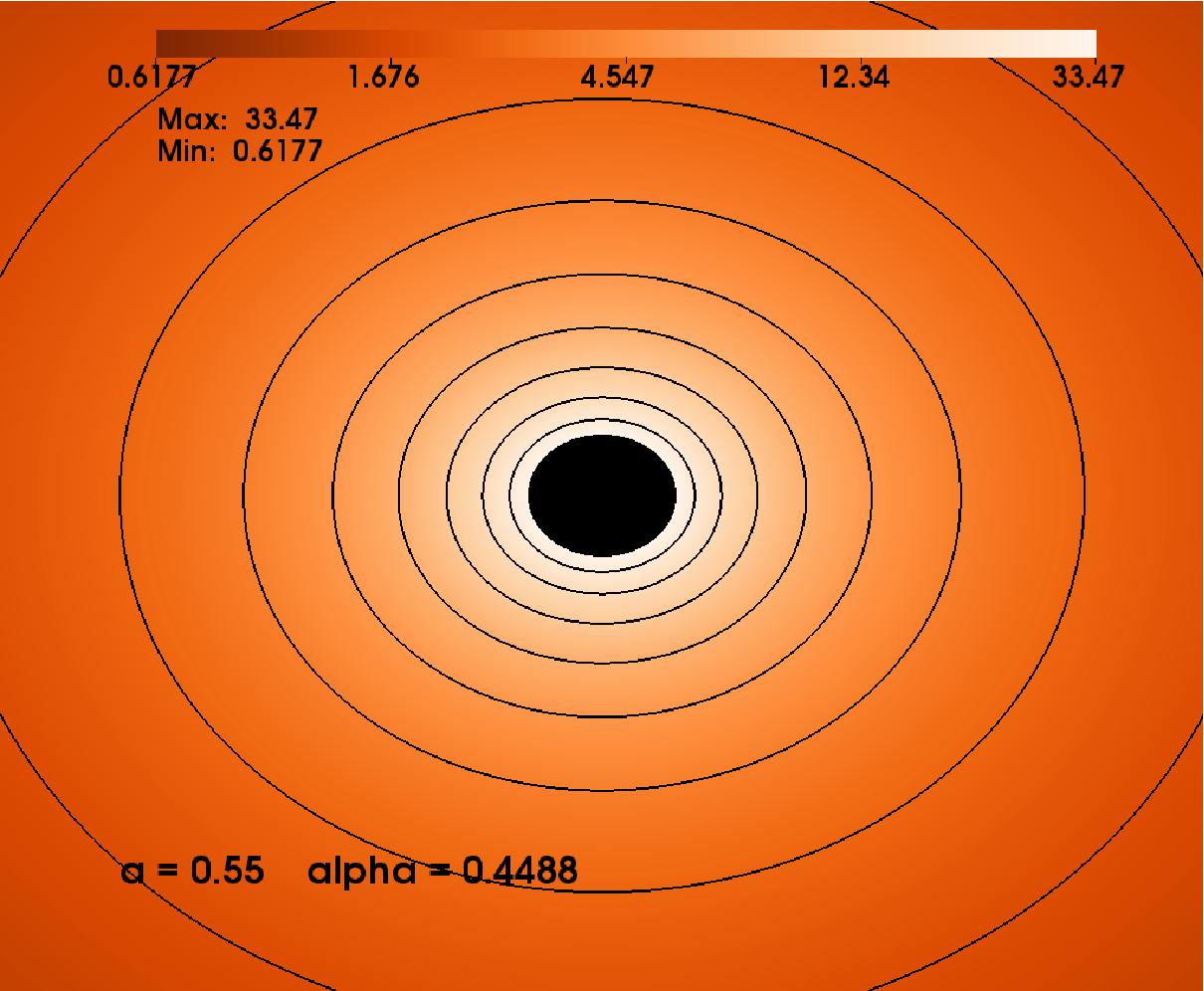,width=7.5cm}
  \psfig{file=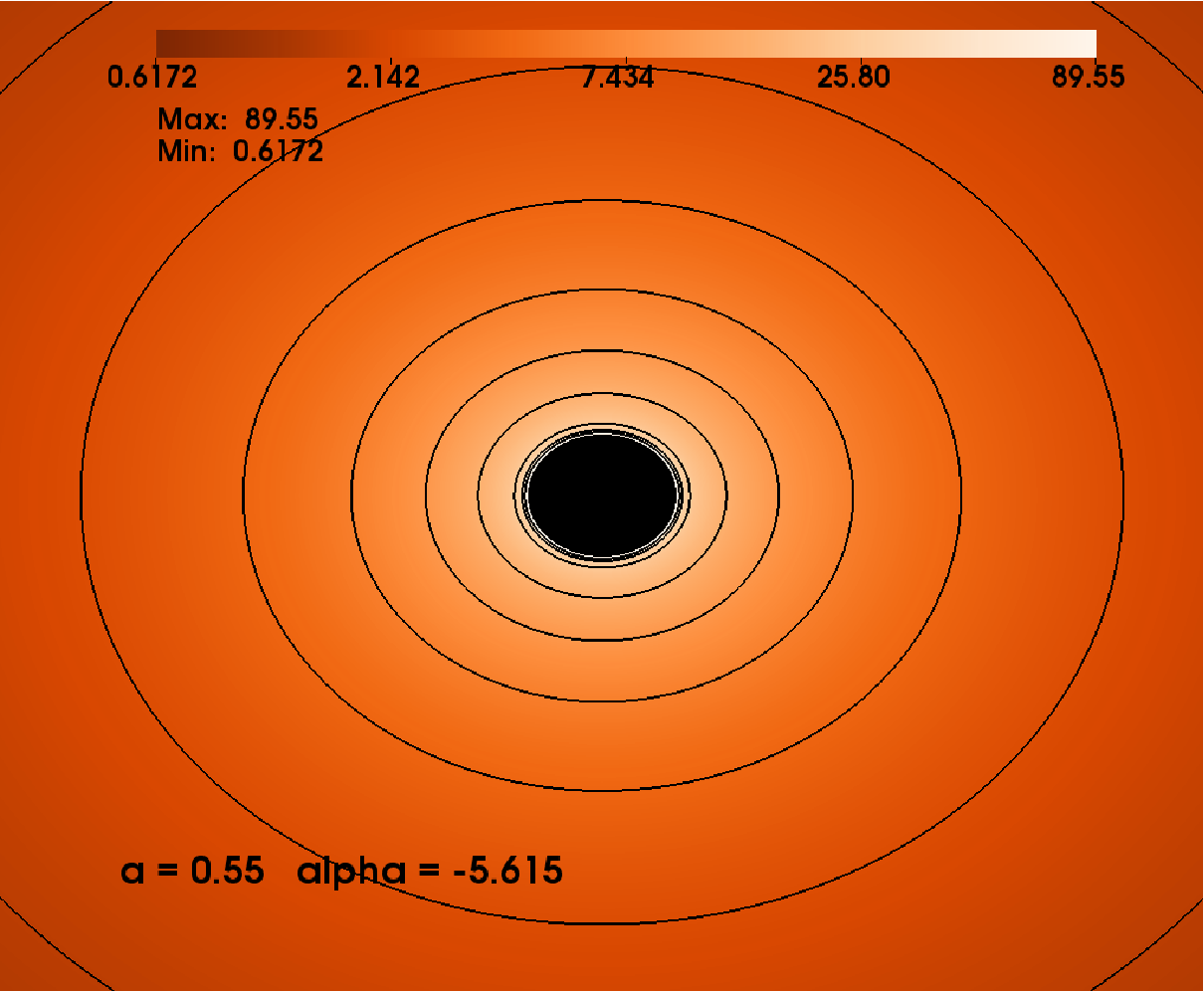,width=7.5cm}
  \psfig{file=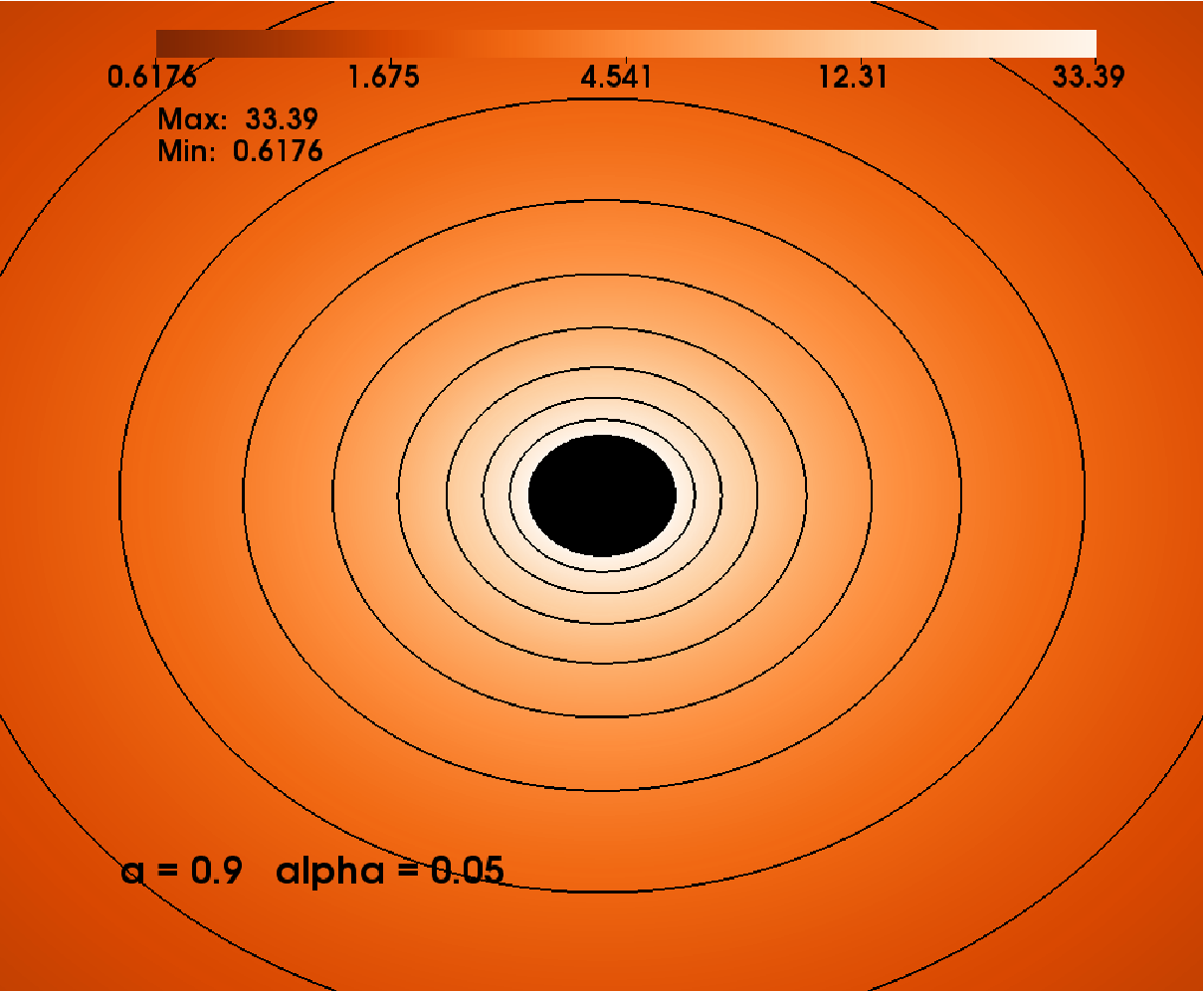,width=7.5cm}
  \psfig{file=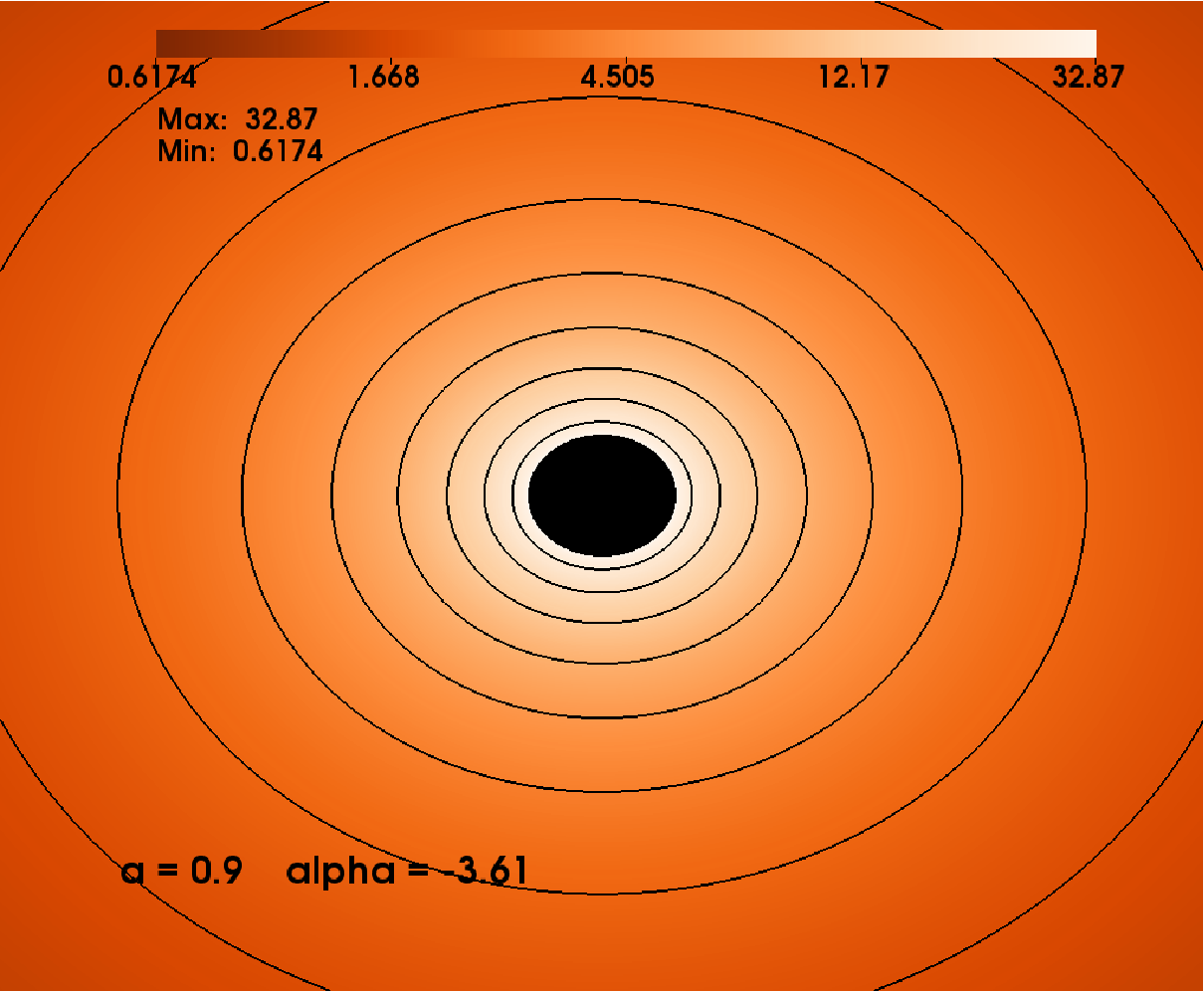,width=7.5cm}
  \caption{The plot displays the rest-mass density using a logarithmic color scale
    conducted within the framework of Extended EGB gravity. In all the models,
    a perfect spherical accretion process occurs, resulting in a relatively uniform
    distribution of matter. The logarithmic color scale and contour lines allow for a clear
    visualization of the variations in the rest-mass density across the simulation
    domain, highlighting the accretion process in the EGB gravity framework.}
\label{Color_logdensity}
\end{figure*}

Supernova remnants can potentially fall towards a newly formed black
hole and contribute to the formation of an accretion disk. When a massive
star undergoes a supernova explosion, the outer layers are ejected into space,
leaving behind a compact object, such as a black hole.
The material expelled during the supernova can interact with the surrounding
medium and, under certain conditions, fall back towards the black hole.
If the material falls back towards the black hole in a disk-like configuration,
it can form an accretion disk as see in in Fig.\ref{Color_logdensity}.
Indeed, as the infalling matter within the accretion disk spirals closer to the black hole,
it experiences gravitational forces that cause it to gain kinetic energy and heat up.
This process releases a significant amount of energy, which is closely related to
the amount of matter being accreted toward the black hole.
This accretion disk plays a vital role in the growth and evolution of the
black hole by providing a source of mass and energy.

The Fig.\ref{Color_logdensity} clearly demonstrates that for positive values of the EGB coupling
constant $(\alpha)$ with different values of the black hole rotation parameter $(a)$,
the maximum density of the accretion disk around the black hole is noticeably reduced compared
to the case with negative values of alpha and different values of the black hole
rotation parameter. This numerical observation suggests that the presence of a positive EGB
coupling constant can affect the accretion process, resulting in a reduced accretion
rate toward the black hole. The specific mechanisms responsible for this reduction can
vary, depending on the details of the system and the interplay between the EGB gravity
and the dynamics of the accretion disk. Understanding the impact of the EGB coupling
constant and the black hole rotation parameter on the accretion process is crucial
for comprehending the behavior of matter in the vicinity of black holes and the
overall evolution of accretion systems influenced by modified theories of gravity
\citep{Johannsen2016, Bambi2018}.
In general, a higher accretion rate, where a larger amount of matter
is falling toward the black hole, will result in a more rapid growth of the black
hole's mass. This relationship between accretion rate and black hole mass growth
is an important aspect of understanding the evolution and dynamics of black holes
in astrophysical systems \citep{Hopkins2008}.

%%%%%%%%%%%%%%%%%%%%%%%%%%%%%%%%%%%%%%%%%%%%%%%%%%%%%%%%%%%%%%%%%%%%%%

\begin{figure*}
  \vspace{1cm}
  \center
  \psfig{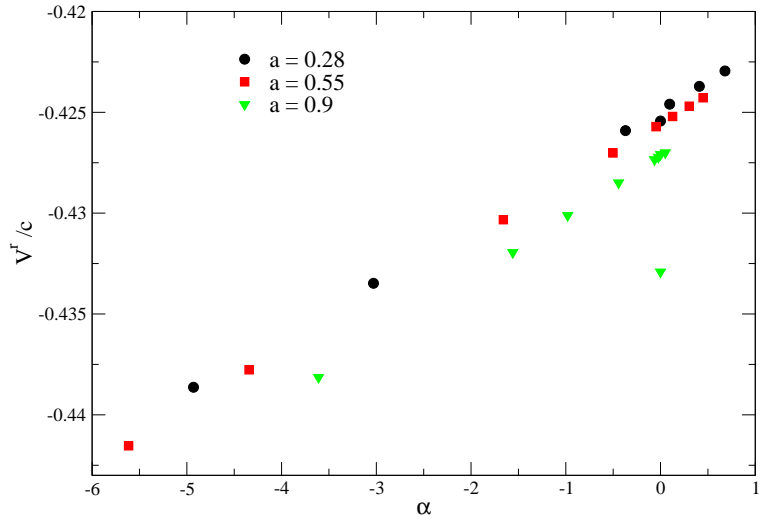}
  \caption{The plot shows the radial infall velocity of matter at a fixed
    position with coordinates $r=5M$ and $\phi=0$ in the strong gravitational
    region at the final time step. The data is presented for different values
    of the EGB coupling constant $(\alpha)$ with different a. The plot provides
    insights into how the infall velocity changes as the EGB coupling constant varies.}
%\vspace{1cm}
\label{VrAlpha}
\end{figure*}

Understanding the infalling velocity of matter toward the black hole provides valuable
insights into various aspects of the accretion process, including the matter's accretion
behavior, its dependence on the EGB coupling constant $(\alpha)$ and black hole
rotation parameter $(a)$, the growth rate
of the black hole mass, and potentially more. By analyzing the infalling velocity,
it is possible to gain information about the dynamics and characteristics of the accretion
process, such as the efficiency of angular momentum transfer, the accretion disk's
structure, and the interaction between the matter and the black hole's gravitational field.
The infalling velocity can provide insights into the accretion rate, which is related
to the amount of matter being accreted toward the black hole and, consequently, the growth
rate of the black hole mass. Fig.\ref{VrAlpha} shows that  by studying how the infalling
velocity varies with different
values of the $\alpha$ parameter (related to the modified theory of gravity) and $a$,
we can investigate how these factors influence the
accretion process and the resulting growth of the black hole.

As the alpha parameter increases in the negative direction seen in Fig.\ref{VrAlpha},
it is observed that the infalling matter velocity becomes more pronounced or accelerated.
This heightened velocity causes a larger
quantity of matter to be drawn towards the black hole.
The increased inflow of matter due to the increased in radial velocity,
as depicted in Figs.\ref{Color_logdensity} and \ref{VrAlpha}, leads to
two significant consequences. Firstly, there is an increase in the maximum
density of the accretion disk.
The greater amount of matter accumulating in the disk results in a higher density
region, reflecting a more concentrated mass distribution around the black hole.
This distribution around a black hole has profound implications for the surrounding
astrophysical environment, influencing the gravitational field, the formation of
accretion disks, and various other phenomena associated with the
black hole's presence \citep{Rees1984}.
Secondly, the heightened infalling matter rate contributes to an increased amount
of material falling directly towards the black hole. This influx of matter toward the
black hole enhances its accretion rate and consequently
accelerates the growth of its mass \citep{Natarajan2004, Mountrichas2023}

On the other hand, the rotation parameter of the spinning black hole,
as shown in Fig.\ref{VrAlpha}, influences the velocity and consequently the amount
of matter falling towards the black hole. A larger rotation parameter
generates a stronger gravitational pull, resulting in faster rotation
of a larger amount of matter.

%%%%%%%%%%%%%%%%%%%%%%%%%%%%%%%%%%%%%%%%%%%%%%%%%%%%%%%%%%%%%%%%%%%%%%

\begin{figure*}
  \vspace{1cm}
  \center
  \psfig{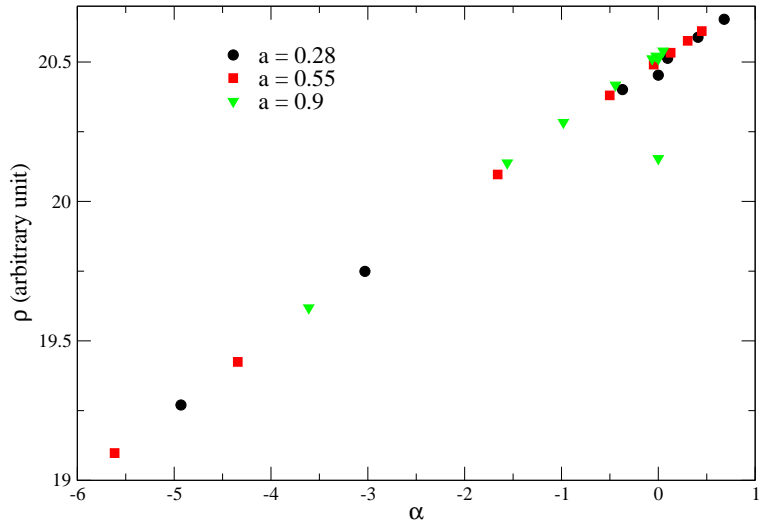}
  \caption{Same as Fig.\ref{VrAlpha} but for the rest mass density of the accreted remnants}
%\vspace{1cm}
\label{densityAlpha}
\end{figure*}

Fig.\ref{densityAlpha}, analogous to Fig.\ref{VrAlpha}, provides insight into the density variation
of the accretion disk surrounding the black hole, depending on the variation of the
$\alpha$ and $a$ parameters. The simulation captures the density distribution
at the final time step, allowing us to discern important trends. Notably, there
is a slight increase in the accreting of matter surrounding the rapidly rotating black hole.
This observed phenomenon aligns with expectations as the curved space-time around the
black hole engenders a higher potential barrier. Consequently, the gravitational pull
exerted on matter is intensified, resulting in a more pronounced attraction towards the
black hole. The accumulation of matter is further amplified by the enhanced gravitational
field, which manifests in a closer proximity of the final stable orbit to the black hole.
The observed increase in density and matter accumulation around the rapidly rotating black
hole signifies the intricate interplay between the curved space-time, stronger
gravitational forces, and the dynamics of the accretion process \citep{Beckwith2005}.

%%%%%%%%%%%%%%%%%%%%%%%%%%%%%%%%%%%%%%%%%%%%%%%%%%%%%%%%%%%%%%%%%%%%%%

The mass accretion rate is a crucial parameter in understanding the dynamics of
an accretion disk and the growth rate of a black hole.
The accretion rate demonstrates the speed at which matter is drawn inward, causing
the disk to heat up. Generally, as accretion rates rise, more energy is produced,
leading to increasingly intricate and high-energy phenomena. Meanwhile, the more
material a black hole accrues during and subsequent to the formation of the
accretion disk, the greater its mass becomes. This growth in the black hole's
mass can influence its inherent properties. Comprehending these characteristics
enables us to glean more detailed insights into the complex structures of
accretion disks, the X-ray emissions surrounding them \citep{Fabian2009},
and gravitational waves\citep{Abbott2021Virgo}.

\begin{figure*}
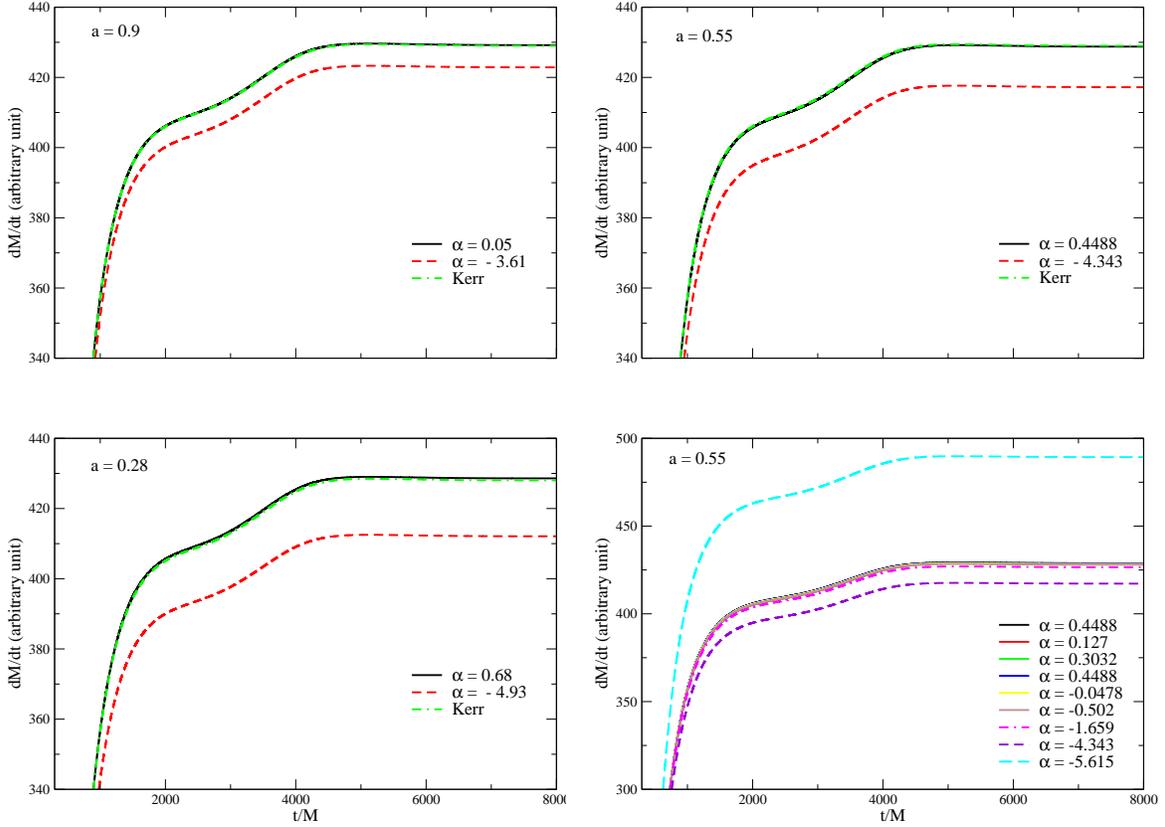

  \vspace{1cm}
  \center
    \psfig{file=Fig4_1.eps,width=7.5cm} \hspace*{0.15cm}
    \psfig{file=Fig4_2.eps,width=7.5cm} \\
     \vspace*{0.5cm} 
    \psfig{file=Fig4_3.eps,width=7.5cm} \hspace*{0.15cm}
    \psfig{file=Fig4_4.eps,width=7.5cm}
    \caption{The graph illustrates the black hole mass accretion rate,
      represented in arbitrary units, as a function of time at the inner
      boundary of the computational domain, located near the black hole
      event horizon. The data is plotted for various values of the black
      hole rotation parameter $(a$) and the EGB coupling constant $(\alpha)$.
      The black straight line corresponds to the case of a Kerr black hole.}
%\vspace{1cm}
\label{Mass_acc}
\end{figure*}

Figure \ref{Mass_acc} presents plots of the mass accretion rates, which have been
calculated based on various values of both the 
EGB coupling constant $(\alpha)$ and the black hole's spin $(a)$.
EGB coupling constant affect the geometry of space-time around the black hole,
and the behavior of the accretion disk, including the mass accretion rate.
The mass accretion rate for positive values of the 
EGB coupling constant appears to exhibit similar patterns to those of a
Kerr black hole. However, when the EGB constant takes on negative values,
the accretion rate is noticeably reduced.
Although the established pattern holds for many values, it doesn't apply when
$\alpha=-5.615$, as depicted in the lower right of Fig.\ref{Mass_acc}. It seems
that when the $\alpha$  reaches more extreme negative values (exceeding -5),
discrepancies arise between the numerical simulations of the accretion disk
and its resulting events, and the analytic solutions and observational data.
To validate this assertion, it's essential to perform additional simulations.

The spin of the black hole is another crucial factor that influences the structure
of the accretion disk and the accretion rate, as it can cause the inner edge of the
accretion disk to move closer to the black hole, leading to more efficient accretion.
But no substantial variations in the accretion rate are observed across different
values of the spin parameter as seen in Fig.\ref{Mass_acc}.

%%%%%%%%%%%%%%%%%%%%%%%%%%%%%%%%%%%%%%%%%%%%%%%%%%%%%%%%%%%%%%%%%%%%%%%

\begin{figure*}
  \vspace{1cm}
  \center
  \psfig{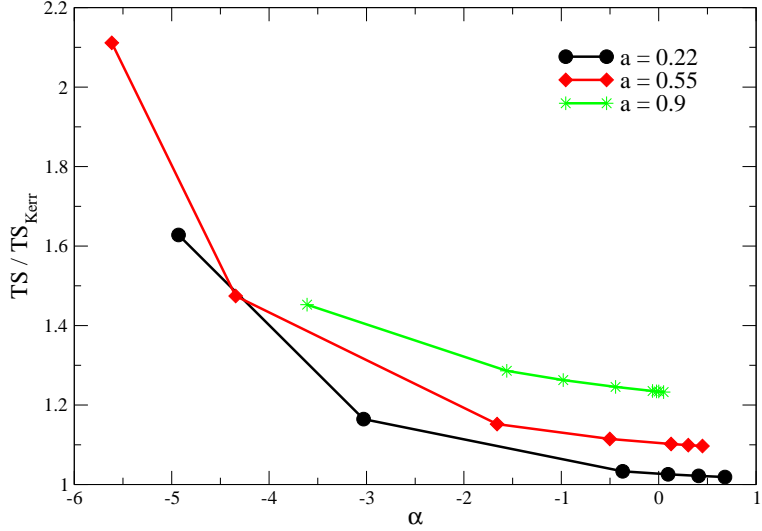}
  \caption{The required numbers of time steps $TS$ which is normalized to the Kerr
    black hole time steps $TS_{Kerr}$  versus EGB constant $\alpha$ is plotted for the different
    values of rotation parameter $(a)$.  The total number of time steps have been reached
    in the maximum time 30000M.}
%\vspace{1cm}
\label{Required_time}
\end{figure*}

As seen in Fig.\ref{Required_time}, to investigate the chaotic
dynamics of the accretion disk mechanism,
we graph the number of time steps from each simulation as a function of
$\alpha$ value. We discover that as the $\alpha$ value becomes increasingly negative,
the structure and interactions within the system become more intricate and complex.
Consequently, numerical simulations involving these  $\alpha$ values demand more
computational resources and time to reach the maximum time $(t_{max}=30000M)$
applied across all simulations.

As anticipated and seen in Fig.\ref{Required_time}, given the intricate dynamics of
accretion when the black hole rotation parameter increases, the number of time
steps also elevates correspondingly with larger values of the black hole spin
parameter. This is due to the complex and rich structure introduced by
higher rotational parameter. The spin of a black hole can profoundly affect the
dynamics and mechanisms of its adjacent accretion disk in numerous intricate ways.
For instance, when the rotation parameters increase, the disk is drawn closer to the
event horizon, enhancing the efficiency of accretion. Furthermore, the spinning
black hole has the capacity to warp surrounding space-time, among other complexities.

%%%%%%%%%%%%%%%%%%%%%%%%%%%%%%%%%%%%%%%%%%%%%%%%%%%%%%%%%%%%%%%%%%%%%%%
\section{The disk instability and QPOs}
\label{QPO}

Non-axisymmetric instabilities that occur within the black hole-accretion
disk system are significant for understanding the behavior of the accreted
matter. The characterization of these instabilities involves studying the
system and determining the associated azimuthal wave number (m) that represents
the specific instability.
To further characterize the instability, the saturation point is determined
through computation using simulation data. This computation involves calculating
the Fourier power in density, which enables the identification of the point at
which the instability reaches a stable or saturated state. It is worth noting
that the formulation and application of the Fourier transform have been
extensively studied by \citet{Donmez4}.
By employing these methods, a more comprehensive occurrence of
the non-axisymmetric instabilities in the black hole-accretion disk
system can be studied. This work leads to valuable insights into the
underlying dynamics and behavior of the system, particularly regarding
the accretion process of matter toward the black hole.

\begin{figure*}
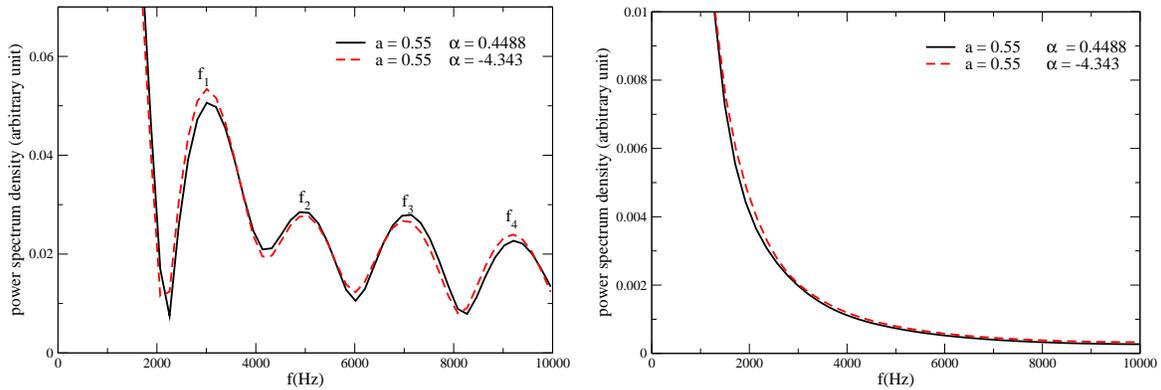

  \vspace{1cm}
  \center
    \psfig{file=Fig6_1.eps,width=7.5cm} \hspace*{0.15cm}
    \psfig{file=Fig6_2.eps,width=7.5cm} \\
    \caption{The power spectrum of the rest mass density is shown for a black hole
      with a mass of $M = 10M_{\odot}$. The plot on the left represents the
      spectrum throughout the entire simulation, capturing the variations in density
      at different frequencies. On the right, the plot displays the power spectrum
      specifically after the disk has reached a steady state.}
%\vspace{1cm}
\label{PowerSpectrum}
\end{figure*}

A power spectrum of the selected model, as shown in Fig.\ref{PowerSpectrum},
is calculated to examine whether there are any oscillations in the rest-mass
density of the accretion disk after it reaches a steady state. It enables a
targeted investigation of the density fluctuations and characteristic
frequencies within the system once it has stabilized. However, the analysis reveals
no observable frequencies, indicating the absence of oscillations in the density of
the accretion disk once it reaches a steady state.

During the initial stages of the accretion disk's formation, the instabilities
manifested around the black hole as shown in the left part of Fig.\ref{PowerSpectrum}.
These instabilities caused the disk to undergo
oscillatory behavior, characterized by regular and repetitive variations in its
properties. These oscillations occurred at specific frequencies, indicating a
pattern in the accretion disk behavior. The presence of these oscillations and their
frequencies provide valuable insights into the dynamic processes taking place
during the early stages of disk formation and can be studied to better understand
the underlying physics governing the system. The oscillations within the black
hole-disk system give rise to quasi-periodic behavior, and the frequency of
quasi-periodic oscillations (QPOs) may be linked to the Keplerian frequency
of the accretion flow at the inner radius around a Kerr black hole \citep{Kumar2023}.

The left panel of Fig.\ref{PowerSpectrum}, displaying the spectra of density,
reveals that the observed oscillations are largely unaffected by changes in the
EGB coupling constant $\alpha$. This suggests that the oscillatory behavior is
independent of this parameter and likely driven by other factors.
During the formation of the accretion disk, global genuine eigenmodes are
detected, $f_1=3000 Hz$, $f_2=5000 Hz$, $f_3=7000 Hz$, and $f_4=9200 Hz$.
These eigenmodes represent collective oscillations that involve the entire system,
indicating coherent behavior within the disk. Interestingly, no non-linear coupling
modes are observed during the evolution of the accretion disk. This implies that the
interactions and dynamics within the system do not exhibit significant non-linear
effects, and the behavior can be largely described by linear processes.

%%%%%%%%%%%%%%%%%%%%%%%%%%%%%%%%%%%%%%%%%%%%%%%%%%%%%%%%%%%%%%%%%%%%%%%
%%%%%%%%%%%%%%%%%%%%%%%%%%%%%%%%%%%%%%%%%%%%%%%%%%%%%%%%%%%%%%%%%%%%%%%

\section{Discussion and Conclusion}
\label{Conclusion}
%%%%%%%%%%%%%%%%%%%%%%%%%%%%%%%%%%%%%%%%%%%%%%%%%%%%%%%%%%%%%%%%%%%%%%%

The formation of an accretion disk from supernova remnants falling toward a
newly formed black hole is a fascinating phenomenon that contributes to our
understanding of the dynamics of black hole formation and subsequent
accretion processes. The formation of a new accretion disk can instigate
radiation emissions, including X-rays. These emissions can be observed and
utilized to uncover the properties of the black hole. Specifically, the
intensity, spectrum, and variability of the X-ray emissions can provide
valuable insights into the mass, spin, and environment of the black hole,
as well as the physics of the accretion process itself.

In this paper, we aimed to investigate how the structure of the accretion
disk, formed by the supernova remnants falling towards the black hole,
depends on the Einstein-Gauss-Bonnet (EGB) and black hole spin parameters.
We found the following results numerically.

\begin{enumerate}[label=\roman*.]
\item We explicitly show that when the EGB coupling constant $(\alpha)$ takes
positive values across various black hole rotation parameters $(a)$,
the maximum density of the accretion disk surrounding the black hole is
substantially lower compared to cases with negative alpha values and
different black hole rotation parameters. This numerical observation implies
that the existence of a positive EGB coupling constant can influence the
accretion process, leading to a diminished accretion
rate toward the black hole \citep{Long2022}.

\item We undertake a detailed examination of how the infalling velocity
  fluctuates with different values of the $\alpha$ parameter and the
  rotation parameter $a$. The space-time geometry around the black hole and
  the dynamics of the accretion disk can be influenced by the EGB constant.
  Consequently, this influence may affect the infalling velocity,
  especially when compared with a Kerr black hole \citep{Liu2021}.
  Additionally, we explore how these variables
  impact the accretion process and the consequent expansion of the black hole.

\item It is observed that the velocity of the matter falling into the black
  hole becomes more substantial or accelerated while  the alpha parameter increases
  towards the negative side. The negative $\alpha$ values are also important to explain
  the astrophysical phenomena \citep{Dehghani2004}.
  The spin parameter of a rotating black hole
  affects the velocity and subsequently the volume of matter being drawn
  towards the black hole. A larger spin parameter creates a more potent
  gravitational attraction, leading to the quicker rotation of a greater mass of matter.

 \item EGB coupling constant influences the geometry of the space-time surrounding
  the black hole and the dynamics of the accretion disk, including the mass accretion rate.
  When the EGB constant is positive, the mass accretion rate seems to follow
  patterns similar to those of a Kerr black hole. However,
  with negative values of the EGB constant,
  there's a noticeable reduction in the accretion rate and it is also observed by \citet{Maurya2022}

 \item  The power spectrum of density reveals that instabilities appear around
  the black hole during the early stages of the accretion disk's formation.
  These instabilities may play a significant role in shaping the dynamics of
  the accretion process. The observed oscillations appear to be largely
  unchanged by alterations in the EGB coupling constant $\alpha$.
  This suggests that these oscillations may be a fundamental characteristic
  of accretion disk dynamics and may not be heavily influenced
  by the EGB coupling constant. Conversely, the analysis shows no discernible
  frequencies, suggesting that once the accretion disk reaches a stable state,
  oscillations in its density are absent. This could imply a shift from chaotic
  to more orderly dynamics as the system evolves towards equilibrium. 
\end{enumerate}

In the future works, the reasons behind the distinct behavior of the accretion
rate for higher negative values of EGB coupling constant $\alpha$ will be explored. Additionally,
the effects of perturbations that may arise after the accretion disk reaches
a steady state around the black hole, and the role of EGB coupling constant $\alpha$
and the lack hole rotation parameter $a$ in such
scenarios, will be considered in planned future studies

%%%%%%%%%%%%%%%%%%%%%%%%%%%%%%%%%%%%%%%%%%%%%%%%%%%%%%%%%%%%%%%%%%%%%%%%%%%%%%%%%%%%%%%%
%%%%%%%%%%%%%%%%%%%%%%%%%%%%%%%%%%%%%%%%%%%%%%%%%%%%%%%%%%%%%%%%%%%%%%%%%%%%%%%%%%%%%%%

\section*{Acknowledgments}
I would like to express my sincere gratitude to the anonymous referee for their meticulous review and
insightful feedback. Their constructive comments significantly improved the quality of this paper.
All simulations were performed using the Phoenix  High
Performance Computing facility at the American University of the Middle East
(AUM), Kuwait.\\

\bibliography{paper.bib}

\begin{thebibliography}{}
\expandafter\ifx\csname natexlab\endcsname\relax\def\natexlab#1{#1}\fi
\providecommand{\url}[1]{\href{#1}{#1}}
\providecommand{\dodoi}[1]{doi:~\href{http://doi.org/#1}{\nolinkurl{#1}}}
\providecommand{\doeprint}[1]{\href{http://ascl.net/#1}{\nolinkurl{http://ascl.net/#1}}}
\providecommand{\doarXiv}[1]{\href{https://arxiv.org/abs/#1}{\nolinkurl{https://arxiv.org/abs/#1}}}

\bibitem[{{Abbott} {et~al.}(2021){Abbott}, {Abbott}, {Abraham}, {Acernese},
  {Ackley}, {Adams}, {Adams}, {Adhikari}, {Adya}, {Affeldt}, {Agarwal},
  {Agathos}, {Agatsuma}, {Aggarwal}, {Aguiar}, {Aiello}, {Ain}, {Ajith},
  {Aleman}, {Allen}, {Allocca}, {Altin}, {Amato}, {Anand}, {Ananyeva},
  {Anderson}, {Anderson}, {Angelova}, {Ansoldi}, {Antelis}, {Antier}, {Appert},
  {Arai}, {Araya}, {Areeda}, {Ar{\`e}ne}, {Arnaud}, {Aronson}, {Arun}, {Asali},
  {Ashton}, {Aston}, {Astone}, {Aubin}, {Aufmuth}, {AultONeal}, {Austin},
  {Babak}, {Badaracco}, {Bader}, {Bae}, {Baer}, {Bagnasco}, {Bai}, {Baird},
  {Ball}, {Ballardin}, {Ballmer}, {Bals}, {Balsamo}, {Baltus}, {Banagiri},
  {Bankar}, {Bankar}, {Barayoga}, {Barbieri}, {Barish}, {Barker}, {Barneo},
  {Barone}, {Barr}, {Barsotti}, {Barsuglia}, {Barta}, {Bartlett}, {Barton},
  {Bartos}, {Bassiri}, {Basti}, {Bawaj}, {Bayley}, {Baylor}, {Bazzan},
  {B{\'e}csy}, {Bedakihale}, {Bejger}, {Belahcene}, {Benedetto}, {Beniwal},
  {Benjamin}, {Bennett}, {Bentley}, {BenYaala}, {Bergamin}, {Berger},
  {Bernuzzi}, {Berry}, {Bersanetti}, {Bertolini}, {Betzwieser}, {Bhandare},
  {Bhandari}, {Bhattacharjee}, {Bhaumik}, {Bidler}, {Bilenko}, {Billingsley},
  {Birney}, {Birnholtz}, {Biscans}, {Bischi}, {Biscoveanu}, {Bisht}, {Biswas},
  {Bitossi}, {Bizouard}, {Blackburn}, {Blackman}, {Blair}, {Blair}, {Blair},
  {Bobba}, {Bode}, {Boer}, {Bogaert}, {Boldrini}, {Bondu}, {Bonilla},
  {Bonnand}, {Booker}, {Boom}, {Bork}, {Boschi}, {Bose}, {Bose}, {Bossilkov},
  {Boudart}, {Bouffanais}, {Bozzi}, {Bradaschia}, {Brady}, {Bramley}, {Branch},
  {Branchesi}, {Brau}, {Breschi}, {Briant}, {Briggs}, {Brillet}, {Brinkmann},
  {Brockill}, {Brooks}, {Brooks}, {Brown}, {Brunett}, {Bruno}, {Bruntz},
  {Bryant}, {Buikema}, {Bulik}, {Bulten}, {Buonanno}, {Buscicchio}, {Buskulic},
  {Byer}, {Cadonati}, {Caesar}, {Cagnoli}, {Cahillane}, {}, {Bustillo},
  {Callaghan}, {Callister}, {Calloni}, {Camp}, {Canepa}, {Cannavacciuolo},
  {Cannon}, {Cao}, {Cao}, {Capote}, {Carapella}, {Carbognani}, {Carlin},
  {Carney}, {Carpinelli}, {Carullo}, {Carver}, {Diaz}, {Casentini}, {Castaldi},
  {Caudill}, {Cavagli{\`a}}, {Cavalier}, {Cavalieri}, {Cella},
  {Cerd{\'a}-Dur{\'a}n}, {Cesarini}, {Chaibi}, {Chakravarti}, {Champion},
  {Chan}, {Chan}, {Chan}, {Chandra}, {Chanial}, {Chao}, {Charlton}, {Chase},
  {Chassande-Mottin}, {Chatterjee}, {Chaturvedi}, {Chen}, {Chen}, {Chen},
  {Chen}, {Chen}, {Chen}, {Cheng}, {Cheong}, {Cheung}, {Chia}, {Chiadini},
  {Chierici}, {Chincarini}, {Chiofalo}, {Chiummo}, {Cho}, {Cho}, {Choate},
  {Choudhary}, {Choudhary}, {Christensen}, {Chu}, {Chua}, {Chung}, {Ciani},
  {Ciecielag}, {Cie{\'s}lar}, {Cifaldi}, {Ciobanu}, {Ciolfi}, {Cipriano},
  {Cirone}, {Clara}, {Clark}, {Clark}, {Clarke}, {Clearwater}, {Clesse},
  {Cleva}, {Coccia}, {Cohadon}, {Cohen}, {Cohen}, {Colleoni}, {Collette},
  {Colpi}, {Compton}, {Constancio}, {Conti}, {Cooper}, {Corban}, {Corbitt},
  {Cordero-Carri{\'o}n}, {Corezzi}, {Corley}, {Cornish}, {Corre}, {Corsi},
  {Cortese}, {Costa}, {Cotesta}, {Coughlin}, {Coughlin}, {Coulon},
  {Countryman}, {Cousins}, {Couvares}, {Covas}, {Coward}, {Cowart}, {Coyne},
  {Coyne}, {Creighton}, {Creighton}, {Criswell}, {Croquette}, {Crowder},
  {Cudell}, {Cullen}, {Cumming}, {Cummings}, {Cuoco}, {Cury{\l}o}, {Canton},
  {D{\'a}lya}, {Dana}, {DaneshgaranBajastani}, {D'Angelo}, {Danilishin},
  {D'Antonio}, {Danzmann}, {Darsow-Fromm}, {Dasgupta}, {Datrier}, {Dattilo},
  {Dave}, {Davier}, {Davies}, {Davis}, {Daw}, {Dean}, {DeBra}, {Deenadayalan},
  {Degallaix}, {Laurentis}, {Del{\'e}glise}, {Favero}, {Lillo}, {Lillo},
  {Pozzo}, {DeMarchi}, {Matteis}, {D'Emilio}, {Demos}, {Dent}, {Depasse},
  {Pietri}, {Rosa}, {Rossi}, {DeSalvo}, {Simone}, {Dhurandhar}, {D{\'\i}az},
  {Diaz-Ortiz}, {Didio}, {Dietrich}, {Di Fiore}, {Di Fronzo}, {Di Giorgio}, {Di
  Giovanni}, {Di Girolamo}, {Di Lieto}, {Ding}, {Di Pace}, {Di Palma}, {Di
  Renzo}, {Divakarla}, {Dmitriev}, {Doctor}, {D'Onofrio}, {Donovan}, {Dooley},
  {Doravari}, {Dorrington}, {Drago}, {Driggers}, {Drori}, {Du}, {Ducoin},
  {Dupej}, {Durante}, {D'Urso}, {Duverne}, {Dwyer}, {Easter}, {Ebersold},
  {Eddolls}, {Edelman}, {Edo}, {Edy}, {Effler}, {Eichholz}, {Eikenberry},
  {Eisenmann}, {Eisenstein}, {Ejlli}, {Errico}, {Essick}, {Estell{\'e}s},
  {Estevez}, {Etienne}, {Etzel}, {Evans}, {Evans}, {Ewing}, {Ezquiaga},
  {Fafone}, {Fair}, {Fairhurst}, {Fan}, {Farah}, {Farinon}, {Farr}, {Farr},
  {Farrow}, {Fauchon-Jones}, {Favata}, {Fays}, {Fazio}, {Feicht}, {Fejer},
  {Feng}, {Fenyvesi}, {Ferguson}, {Fernandez-Galiana}, {Ferrante}, {Ferreira},
  {Fidecaro}, {Figura}, {Fiori}, {Fishbach}, {Fisher}, {Fittipaldi}, {Fiumara},
  {Flaminio}, {Floden}, {Flynn}, {Fong}, {Font}, {Fornal}, {Forsyth}, {Franke},
  {Frasca}, {Frasconi}, {Frederick}, {Frei}, {Freise}, {Frey}, {Fritschel},
  {Frolov}, {Fronz{\'e}}, {Fulda}, {Fyffe}, {Gabbard}, {Gadre}, {Gaebel},
  {Gair}, {Gais}, {Galaudage}, {Gamba}, {Ganapathy}, {Ganguly}, {Gaonkar},
  {Garaventa}, {Garc{\'\i}a-N{\'u}{\~n}ez}, {Garc{\'\i}a-Quir{\'o}s}, {Garufi},
  {Gateley}, {Gaudio}, {Gayathri}, {Gemme}, {Gennai}, {George}, {Gergely},
  {Gewecke}, {Ghonge}, {Ghosh}, {Ghosh}, {Ghosh}, {Ghosh}, {Ghosh},
  {Giacomazzo}, {Giacoppo}, {Giaime}, {Giardina}, {Gibson}, {Gier}, {Giesler},
  {Giri}, {Gissi}, {Glanzer}, {Gleckl}, {Godwin}, {Goetz}, {Goetz}, {Gohlke},
  {Goncharov}, {Gonz{\'a}lez}, {Gopakumar}, {Gosselin}, {Gouaty}, {Goyal},
  {Grace}, {Grado}, {Granata}, {Granata}, {Grant}, {Gras}, {Grassia}, {Gray},
  {Gray}, {Greco}, {Green}, {Green}, {Gretarsson}, {Gretarsson}, {Griffith},
  {Griffiths}, {Griggs}, {Grignani}, {Grimaldi}, {Grimes}, {Grimm}, {Grote},
  {Grunewald}, {Gruning}, {Guerrero}, {Guidi}, {Guimaraes}, {Guix{\'e}},
  {Gulati}, {Guo}, {Guo}, {Gupta}, {Gupta}, {Gupta}, {Gustafson}, {Gustafson},
  {Guzman}, {Haegel}, {Halim}, {Hall}, {Hamilton}, {Hammond}, {Haney}, {Hanks},
  {Hanna}, {Hannam}, {Hannuksela}, {Hansen}, {Hansen}, {Hanson}, {Harder},
  {Hardwick}, {Haris}, {Harms}, {Harry}, {Harry}, {Hartwig}, {Haskell},
  {Hasskew}, {Haster}, {Haughian}, {Hayes}, {Healy}, {Heidmann}, {Heintze},
  {Heinze}, {Heinzel}, {Heitmann}, {Hellman}, {Hello}, {Helmling-Cornell},
  {Hemming}, {Hendry}, {Heng}, {Hennes}, {Hennig}, {Hennig}, {Vivanco},
  {Heurs}, {Hild}, {Hill}, {Hines}, {Hochheim}, {Hofman}, {Hohmann}, {Holgado},
  {Holland}, {Hollows}, {Holmes}, {Holt}, {Holz}, {Hopkins}, {Hough}, {Howell},
  {Hoy}, {Hoyland}, {Hreibi}, {Hsu}, {Huang}, {H{\"u}bner}, {Huddart},
  {Huerta}, {Hughey}, {Hui}, {Husa}, {Huttner}, {Huxford}, {Huynh-Dinh},
  {Idzkowski}, {Iess}, {Inchauspe}, {Ingram}, {Intini}, {Isi}, {Isleif},
  {Iyer}, {JaberianHamedan}, {Jacqmin}, {Jadhav}, {Jadhav}, {James}, {Jan},
  {Jani}, {Janquart}, {Janssens}, {Janthalur}, {Jaranowski}, {Jariwala},
  {Jaume}, {Jenkins}, {Jeunon}, {Jia}, {Jiang}, {Johns}, {Jones}, {Jones},
  {Jones}, {Jones}, {Jones}, {Jonker}, {Ju}, {Junker}, {Kalaghatgi},
  {Kalogera}, {Kamai}, {Kandhasamy}, {Kang}, {Kanner}, {Kao}, {Kapadia},
  {Kapasi}, {Karat}, {Karathanasis}, {Karki}, {Kashyap}, {Kasprzack},
  {Kastaun}, {Katsanevas}, {Katsavounidis}, {Katzman}, {Kaur}, {Kawabe},
  {K{\'e}f{\'e}lian}, {Keitel}, {Key}, {Khadka}, {Khalili}, {Khan}, {Khan},
  {Khazanov}, {Khetan}, {Khursheed}, {Kijbunchoo}, {Kim}, {Kim}, {Kim}, {Kim},
  {Kim}, {Kimball}, {King}, {Kinley-Hanlon}, {Kirchhoff}, {Kissel},
  {Kleybolte}, {Klimenko}, {Knee}, {Knowles}, {Knyazev}, {Koch}, {Koekoek},
  {Koley}, {Kolitsidou}, {Kolstein}, {Komori}, {Kondrashov}, {Kontos}, {Koper},
  {Korobko}, {Kovalam}, {Kozak}, {Kringel}, {Krishnendu}, {Kr{\'o}lak},
  {Kuehn}, {Kuei}, {Kumar}, {Kumar}, {Kumar}, {Kumar}, {Kuns}, {Kwang},
  {Laghi}, {Lalande}, {Lam}, {Lamberts}, {Landry}, {Lane}, {Lang}, {Lange},
  {Lantz}, {Rosa}, {Lartaux-Vollard}, {Lasky}, {Laxen}, {Lazzarini}, {Lazzaro},
  {Leaci}, {Leavey}, {Lecoeuche}, {Lee}, {Lee}, {Lee}, {Lee}, {Lehmann},
  {Lema{\^\i}tre}, {Leon}, {Leroy}, {Letendre}, {Levin}, {Leviton}, {Li}, {Li},
  {Li}, {Li}, {Li}, {Linde}, {Linker}, {Linley}, {Littenberg}, {Liu}, {Liu},
  {Liu}, {Llorens-Monteagudo}, {Lo}, {Lockwood}, {Lollie}, {London}, {Longo},
  {Lopez}, {Lorenzini}, {Loriette}, {Lormand}, {Losurdo}, {Lough}, {Lousto},
  {Lovelace}, {L{\"u}ck}, {Lumaca}, {Lundgren}, {Macas}, {MacInnis}, {Macleod},
  {MacMillan}, {Macquet}, {Hernandez}, {Maga{\~n}a-Sandoval}, {Magazz{\`u}},
  {Magee}, {Maggiore}, {Majorana}, {Makarem}, {Maksimovic}, {Maliakal},
  {Malik}, {Man}, {Mandic}, {Mangano}, {Mango}, {Mansell}, {Manske},
  {Mantovani}, {Mapelli}, {Marchesoni}, {Marion}, {Mark}, {M{\'a}rka},
  {M{\'a}rka}, {Markakis}, {Markosyan}, {Markowitz}, {Maros}, {Marquina},
  {Marsat}, {Martelli}, {Martin}, {Martin}, {Martinez}, {Martinez},
  {Martinovic}, {Martynov}, {Marx}, {Masalehdan}, {Mason}, {Massera},
  {Masserot}, {Massinger}, {Masso-Reid}, {Mastrogiovanni}, {Matas},
  {Mateu-Lucena}, {Matichard}, {Matiushechkina}, {Mavalvala}, {McCann},
  {McCarthy}, {McClelland}, {McClincy}, {McCormick}, {McCuller}, {McGhee},
  {McGuire}, {McIsaac}, {McIver}, {McManus}, {McRae}, {McWilliams}, {Meacher},
  {Mehmet}, {Mehta}, {Melatos}, {Melchor}, {Mendell}, {Menendez-Vazquez},
  {Menoni}, {Mercer}, {Mereni}, {Merfeld}, {Merilh}, {Merritt}, {Merzougui},
  {Meshkov}, {Messenger}, {Messick}, {Meyers}, {Meylahn}, {Mhaske}, {Miani},
  {Miao}, {Michaloliakos}, {Michel}, {Middleton}, {Milano}, {Miller},
  {Millhouse}, {Mills}, {Milotti}, {Milovich-Goff}, {Minazzoli}, {Minenkov},
  {Mir}, {Mishkin}, {Mishra}, {Mishra}, {Mistry}, {Mitra}, {Mitrofanov},
  {Mitselmakher}, {Mittleman}, {Mo}, {Mogushi}, {Mohapatra}, {Mohite},
  {Molina}, {Molina-Ruiz}, {Mondin}, {Montani}, {Moore}, {Moraru}, {Morawski},
  {More}, {Moreno}, {Moreno}, {Morisaki}, {Mours}, {Mow-Lowry}, {Mozzon},
  {Muciaccia}, {Mukherjee}, {Mukherjee}, {Mukherjee}, {Mukherjee}, {Mukund},
  {Mullavey}, {Munch}, {Mu{\~n}iz}, {Murray}, {Musenich}, {Nadji}, {Nagar},
  {Nardecchia}, {Naticchioni}, {Nayak}, {Nayak}, {Neil}, {Neilson}, {Nelemans},
  {Nelson}, {Nery}, {Neunzert}, {Ng}, {Ng}, {Nguyen}, {Nguyen}, {Nguyen},
  {Nichols}, {Nissanke}, {Nocera}, {Noh}, {Norman}, {North}, {Nuttall},
  {Oberling}, {O'Brien}, {O'Dell}, {Oganesyan}, {Oh}, {Oh}, {Ohme}, {Ohta},
  {Okada}, {Olivetto}, {Oram}, {O'Reilly}, {Ormiston}, {Ormsby}, {Ortega},
  {O'Shaughnessy}, {O'Shea}, {Ossokine}, {Osthelder}, {Ottaway}, {Overmier},
  {Pace}, {Pagano}, {Page}, {Pagliaroli}, {Pai}, {Pai}, {Palamos}, {Palashov},
  {Palomba}, {Panda}, {Pang}, {Pankow}, {Pannarale}, {Pant}, {Paoletti},
  {Paoli}, {Paolone}, {Parker}, {Pascucci}, {Pasqualetti}, {Passaquieti},
  {Passuello}, {Patel}, {Patricelli}, {Payne}, {Pechsiri}, {Pedraza},
  {Pegoraro}, {Pele}, {Penn}, {Perego}, {Pereira}, {Pereira}, {Perez},
  {P{\'e}rigois}, {Perreca}, {Perri{\`e}s}, {Petermann}, {Petterson},
  {Pfeiffer}, {Pham}, {Phukon}, {Piccinni}, {Pichot}, {Piendibene},
  {Piergiovanni}, {Pierini}, {Pierro}, {Pillant}, {Pilo}, {Pinard}, {Pinto},
  {Piotrzkowski}, {Piotrzkowski}, {Pirello}, {Pitkin}, {Placidi}, {Plastino},
  {Pluchar}, {Poggiani}, {Polini}, {Pong}, {Ponrathnam}, {Popolizio}, {Porter},
  {Powell}, {Pracchia}, {Pradier}, {Prajapati}, {Prasai}, {Prasanna},
  {Pratten}, {Prestegard}, {Principe}, {Prodi}, {Prokhorov}, {Prosposito},
  {Prudenzi}, {Puecher}, {Punturo}, {Puosi}, {Puppo}, {P{\"u}rrer}, {Qi},
  {Quetschke}, {Quinonez}, {Quitzow-James}, {Raab}, {Raaijmakers}, {Radkins},
  {Radulesco}, {Raffai}, {Rail}, {Raja}, {Rajan}, {Ramirez}, {Ramirez},
  {Ramos-Buades}, {Rana}, {Rapagnani}, {Rapol}, {Ratto}, {Raymond}, {Raza},
  {Razzano}, {Read}, {Rees}, {Regimbau}, {Rei}, {Reid}, {Reitze}, {Relton},
  {Rettegno}, {Ricci}, {Richardson}, {Richardson}, {Richardson}, {Ricker},
  {Riemenschneider}, {Riles}, {Rizzo}, {Robertson}, {Robie}, {Robinet},
  {Rocchi}, {Rocha}, {Rodriguez}, {Rodriguez-Soto}, {Rolland}, {Rollins},
  {Roma}, {Romanelli}, {Romano}, {Romel}, {Romero}, {Romero-Shaw}, {Romie},
  {Rose}, {Rosi{\'n}ska}, {Rosofsky}, {Ross}, {Rowan}, {Rowlinson}, {Roy},
  {Roy}, {Rozza}, {Ruggi}, {Ryan}, {Sachdev}, {Sadecki}, {Sadiq},
  {Sakellariadou}, {Salafia}, {Salconi}, {Saleem}, {Salemi}, {Samajdar},
  {Sanchez}, {Sanchez}, {Sanchez}, {Sanchis-Gual}, {Sanders}, {Sanuy},
  {Saravanan}, {Sarin}, {Sassolas}, {Satari}, {Sathyaprakash}, {Sauter},
  {Savage}, {Savant}, {Sawant}, {Sawant}, {Sayah}, {Schaetzl}, {Scheel},
  {Scheuer}, {Schindler-Tyka}, {Schmidt}, {Schnabel}, {Schneewind},
  {Schofield}, {Sch{\"o}nbeck}, {Schulte}, {Schutz}, {Schwartz}, {Scott},
  {Scott}, {Seglar-Arroyo}, {Seidel}, {Sellers}, {Sengupta}, {Sennett},
  {Sentenac}, {Seo}, {Sequino}, {Sergeev}, {Setyawati}, {Shaffer}, {Shahriar},
  {Shams}, {Sharifi}, {Sharma}, {Sharma}, {Shawhan}, {Shcheblanov}, {Shen},
  {Shikauchi}, {Shink}, {Shoemaker}, {Shoemaker}, {Shukla}, {ShyamSundar},
  {Sieniawska}, {Sigg}, {Singer}, {Singh}, {Singh}, {Singha}, {Sintes},
  {Sipala}, {Skliris}, {Slagmolen}, {Slaven-Blair}, {Smetana}, {Smith},
  {Smith}, {Somala}, {Son}, {Soni}, {Soni}, {Sorazu}, {Sordini}, {Sorrentino},
  {Sorrentino}, {Soulard}, {Souradeep}, {Sowell}, {Spagnuolo}, {Spencer},
  {Spera}, {Srivastava}, {Srivastava}, {Staats}, {Stachie}, {Steer},
  {Steinlechner}, {Steinlechner}, {Stops}, {Stover}, {Strain}, {Strang},
  {Stratta}, {Strunk}, {Sturani}, {Stuver}, {S{\"u}dbeck}, {Sudhagar},
  {Sudhir}, {Suh}, {Summerscales}, {Sun}, {Sun}, {Sunil}, {Sur}, {Suresh},
  {Sutton}, {Swinkels}, {Szczepa{\'n}czyk}, {Szewczyk}, {Tacca}, {Tait},
  {Talbot}, {Tanasijczuk}, {Tanner}, {Tao}, {Tapia}, {San Martin}, {Tasson},
  {Tenorio}, {Terkowski}, {Test}, {Thirugnanasambandam}, {Thomas}, {Thomas},
  {Thompson}, {Thondapu}, {Thorne}, {Thrane}, {Tiwari}, {Tiwari}, {Tiwari},
  {Toland}, {Tolley}, {Tonelli}, {Torres-Forn{\'e}}, {Torrie}, {e Melo},
  {T{\"o}yr{\"a}}, {Trapananti}, {Travasso}, {Traylor}, {Tringali},
  {Tripathee}, {Troiano}, {Trovato}, {Trudeau}, {Tsai}, {Tsai}, {Tsang}, {Tse},
  {Tso}, {Tsukada}, {Tsuna}, {Tsutsui}, {Turconi}, {Ubhi}, {Udall}, {Ueno},
  {Ugolini}, {Unnikrishnan}, {Urban}, {Usman}, {Utina}, {Vahlbruch}, {Vajente},
  {Vajpeyi}, {Valdes}, {Valentini}, {Valsan}, {van Bakel}, {van Beuzekom}, {van
  den Brand}, {Van Den Broeck}, {Vander-Hyde}, {van der Schaaf}, {van
  Heijningen}, {Vanosky}, {Vardaro}, {Vargas}, {Varma}, {Vas{\'u}th},
  {Vecchio}, {Vedovato}, {Veitch}, {Veitch}, {Venkateswara}, {Venneberg},
  {Venugopalan}, {Verkindt}, {Verma}, {Veske}, {Vetrano}, {Vicer{\'e}},
  {Viets}, {Villa-Ortega}, {Vinet}, {Vitale}, {Vo}, {Vocca}, {von Reis}, {von
  Wrangel}, {Vorvick}, {Vyatchanin}, {Wade}, {Wade}, {Wagner}, {Walet},
  {Walker}, {Wallace}, {Wallace}, {Walsh}, {Wang}, {Wang}, {Ward}, {Warner},
  {Was}, {Washington}, {Watchi}, {Weaver}, {Wei}, {Weinert}, {Weinstein},
  {Weiss}, {Weller}, {Wellmann}, {Wen}, {We{\ss}els}, {Westhouse}, {Wette},
  {Whelan}, {White}, {Whiting}, {Whittle}, {Wilken}, {Williams}, {Williams},
  {Williamson}, {Willis}, {Willke}, {Wilson}, {Winkler}, {Wipf}, {Wlodarczyk},
  {Woan}, {Woehler}, {Wofford}, {Wong}, {Wright}, {Wu}, {Wysocki}, {Xiao},
  {Yamamoto}, {Yang}, {Yang}, {Yang}, {Yang}, {Yap}, {Yeeles}, {Yelikar},
  {Yeung}, {Ying}, {Yoon}, {Yu}, {Yu}, {Zadro{\.z}ny}, {Zanolin}, {Zelenova},
  {Zendri}, {Zevin}, {Zhang}, {Zhang}, {Zhang}, {Zhang}, {Zhao}, {Zhao},
  {Zhao}, {Zhou}, {Zhu}, {Zimmerman}, {Zucker}, {Zweizig}, {LIGO Scientific
  Collaboration}, \& {Virgo Collaboration}}]{Abbott2021Virgo}
{Abbott}, R., {Abbott}, T.~D., {Abraham}, S., {et~al.} 2021, \apj, 923, 14,
  \dodoi{10.3847/1538-4357/ac23db}

\bibitem[{{Abramowicz} \& {Fragile}(2013)}]{Abramowicz2013}
{Abramowicz}, M.~A., \& {Fragile}, P.~C. 2013, Living Reviews in Relativity,
  16, 1, \dodoi{10.12942/lrr-2013-1}

\bibitem[{Alcubierre(2008)}]{Alcubierre10}
Alcubierre, M. 2008, in {Introduction to 3+1 Numerical Relativity} (Oxford
  University Press), \dodoi{10.1093/acprof:oso/9780199205677.003.0006}

\bibitem[{{Arrechea} {et~al.}(2020){Arrechea}, {Delhom}, \&
  {Jim{\'e}nez-Cano}}]{Arrechea2020PhRvL}
{Arrechea}, J., {Delhom}, A., \& {Jim{\'e}nez-Cano}, A. 2020, \prl, 125,
  149002, \dodoi{10.1103/PhysRevLett.125.149002}

\bibitem[{{Bambi}(2018)}]{Bambi2018}
{Bambi}, C. 2018, Annalen der Physik, 530, 1700430,
  \dodoi{10.1002/andp.201700430}

\bibitem[{{Banerjee} {et~al.}(2021){Banerjee}, {Tangphati}, \&
  {Channuie}}]{Banerjee2021ApJ}
{Banerjee}, A., {Tangphati}, T., \& {Channuie}, P. 2021, \apj, 909, 14,
  \dodoi{10.3847/1538-4357/abd094}

\bibitem[{{Beckwith} \& {Done}(2005)}]{Beckwith2005}
{Beckwith}, K., \& {Done}, C. 2005, \apss, 300, 87,
  \dodoi{10.1007/s10509-005-1208-5}

\bibitem[{{Bonifacio} {et~al.}(2020){Bonifacio}, {Hinterbichler}, \&
  {Johnson}}]{Bonifacio2020PhRvD}
{Bonifacio}, J., {Hinterbichler}, K., \& {Johnson}, L.~A. 2020, \prd, 102,
  024029, \dodoi{10.1103/PhysRevD.102.024029}

\bibitem[{{Bournaud} {et~al.}(2012){Bournaud}, {Juneau}, {Le Floc'h},
  {Mullaney}, {Daddi}, {Dekel}, {Duc}, {Elbaz}, {Salmi}, \&
  {Dickinson}}]{Bournaud2012}
{Bournaud}, F., {Juneau}, S., {Le Floc'h}, E., {et~al.} 2012, \apj, 757, 81,
  \dodoi{10.1088/0004-637X/757/1/81}

\bibitem[{{Casalino} {et~al.}(2021){Casalino}, {Coll{\'e}aux}, {Rinaldi}, \&
  {Vicentini}}]{Casalino2021PDU}
{Casalino}, A., {Coll{\'e}aux}, A., {Rinaldi}, M., \& {Vicentini}, S. 2021,
  Physics of the Dark Universe, 31, 100770, \dodoi{10.1016/j.dark.2020.100770}

\bibitem[{{Ciotti} \& {Ostriker}(2007)}]{Ciotti2007}
{Ciotti}, L., \& {Ostriker}, J.~P. 2007, \apj, 665, 1038,
  \dodoi{10.1086/519833}

\bibitem[{{Ciotti} \& {Ostriker}(2012)}]{Ciotti2012}
{Ciotti}, L., \& {Ostriker}, J.~P. 2012, in Astrophysics and Space Science
  Library, Vol. 378, Astrophysics and Space Science Library, ed. D.-W. {Kim} \&
  S.~{Pellegrini}, 83, \dodoi{10.1007/978-1-4614-0580-1_4}

\bibitem[{{de Jong} {et~al.}(2022){de Jong}, {Aurrekoetxea}, \&
  {Lim}}]{Jong2022JCAP}
{de Jong}, E., {Aurrekoetxea}, J.~C., \& {Lim}, E.~A. 2022, \jcap, 2022, 029,
  \dodoi{10.1088/1475-7516/2022/03/029}

\bibitem[{{DeGraf} \& {Sijacki}(2017)}]{DeGraf2017}
{DeGraf}, C., \& {Sijacki}, D. 2017, \mnras, 466, 3331,
  \dodoi{10.1093/mnras/stw3267}

\bibitem[{{Dehghani}(2004)}]{Dehghani2004}
{Dehghani}, M.~H. 2004, \prd, 70, 064019, \dodoi{10.1103/PhysRevD.70.064019}

\bibitem[{{D{\"o}nmez}(2004)}]{Donmez1}
{D{\"o}nmez}, O. 2004, \apss, 293, 323,
  \dodoi{10.1023/B:ASTR.0000044610.53714.95}

\bibitem[{{Donmez}(2006)}]{Donmez2}
{Donmez}, O. 2006, AM\&C, 181, 256, \dodoi{10.1016/j.amc.2006.01.031}

\bibitem[{{D{\"o}nmez}(2012)}]{Donmez5}
{D{\"o}nmez}, O. 2012, \mnras, 426, 1533,
  \dodoi{10.1111/j.1365-2966.2012.21616.x}

\bibitem[{{D{\"o}nmez}(2014)}]{Donmez4}
---. 2014, \mnras, 438, 846, \dodoi{10.1093/mnras/stt2255}

\bibitem[{{Donmez}(2022)}]{Donmez_EGB_Rot}
{Donmez}, O. 2022, Physics Letters B, 827, 136997,
  \dodoi{10.1016/j.physletb.2022.136997}

\bibitem[{{Donmez} {et~al.}(2022){Donmez}, {Dogan}, \&
  {Sahin}}]{Donmezetal2022}
{Donmez}, O., {Dogan}, F., \& {Sahin}, T. 2022, Universe, 8, 458,
  \dodoi{10.3390/universe8090458}

\bibitem[{{D{\"o}nmez} {et~al.}(2011){D{\"o}nmez}, {Zanotti}, \&
  {Rezzolla}}]{Donmez6}
{D{\"o}nmez}, O., {Zanotti}, O., \& {Rezzolla}, L. 2011, \mnras, 412, 1659,
  \dodoi{10.1111/j.1365-2966.2010.18003.x}

\bibitem[{{Donmez, Orhan}(2021)}]{Donmez3}
{Donmez, Orhan}. 2021, Eur. Phys. J. C, 81, 113,
  \dodoi{10.1140/epjc/s10052-021-08923-1}

\bibitem[{{Fabian} {et~al.}(2009){Fabian}, {Zoghbi}, {Ross}, {Uttley}, {Gallo},
  {Brandt}, {Blustin}, {Boller}, {Caballero-Garcia}, {Larsson}, {Miller},
  {Miniutti}, {Ponti}, {Reis}, {Reynolds}, {Tanaka}, \& {Young}}]{Fabian2009}
{Fabian}, A.~C., {Zoghbi}, A., {Ross}, R.~R., {et~al.} 2009, \nat, 459, 540,
  \dodoi{10.1038/nature08007}

\bibitem[{{Gatti} {et~al.}(2016){Gatti}, {Shankar}, {Bouillot}, {Menci},
  {Lamastra}, {Hirschmann}, \& {Fiore}}]{Gatti2016}
{Gatti}, M., {Shankar}, F., {Bouillot}, V., {et~al.} 2016, \mnras, 456, 1073,
  \dodoi{10.1093/mnras/stv2754}

\bibitem[{{Ghosh} \& {Maharaj}(2020)}]{Ghosh2020PDU}
{Ghosh}, S.~G., \& {Maharaj}, S.~D. 2020, Physics of the Dark Universe, 30,
  100687, \dodoi{10.1016/j.dark.2020.100687}

\bibitem[{{Glavan} \& {Lin}(2020)}]{Glavan2020PhRvL}
{Glavan}, D., \& {Lin}, C. 2020, \prl, 124, 081301,
  \dodoi{10.1103/PhysRevLett.124.081301}

\bibitem[{{G{\"u}rses} {et~al.}(2020){G{\"u}rses}, {{\c{S}}i{\c{s}}man}, \&
  {Tekin}}]{Metin2020EPJC}
{G{\"u}rses}, M., {{\c{S}}i{\c{s}}man}, T.~{\c{c}}., \& {Tekin}, B. 2020,
  European Physical Journal C, 80, 647, \dodoi{10.1140/epjc/s10052-020-8200-7}

\bibitem[{{Hennigar} {et~al.}(2020){Hennigar}, {Kubiz{\r{A}}{\'a}k}, {Mann}, \&
  {Pollack}}]{Hennigar2020JHEP}
{Hennigar}, R.~A., {Kubiz{\r{A}}{\'a}k}, D., {Mann}, R.~B., \& {Pollack}, C.
  2020, Journal of High Energy Physics, 2020, 27,
  \dodoi{10.1007/JHEP07(2020)027}

\bibitem[{{Heydari-Fard} {et~al.}(2021){Heydari-Fard}, {Heydari-Fard}, \&
  {Sepangi}}]{Fard1}
{Heydari-Fard}, M., {Heydari-Fard}, M., \& {Sepangi}, H.~R. 2021, European
  Physical Journal C, 81, 473, \dodoi{10.1140/epjc/s10052-021-09266-7}

\bibitem[{{Heydari-Fard} \& {Sepangi}(2021)}]{Haydari2}
{Heydari-Fard}, M., \& {Sepangi}, H.~R. 2021, Physics Letters B, 816, 136276,
  \dodoi{10.1016/j.physletb.2021.136276}

\bibitem[{{Hopkins} {et~al.}(2008){Hopkins}, {Cox}, {Kere{\v{s}}}, \&
  {Hernquist}}]{Hopkins2008}
{Hopkins}, P.~F., {Cox}, T.~J., {Kere{\v{s}}}, D., \& {Hernquist}, L. 2008,
  \apjs, 175, 390, \dodoi{10.1086/524363}

\bibitem[{{Hopkins} \& {Hernquist}(2006)}]{Hopkins2006}
{Hopkins}, P.~F., \& {Hernquist}, L. 2006, \apjs, 166, 1,
  \dodoi{10.1086/505753}

\bibitem[{{Johannsen} {et~al.}(2016){Johannsen}, {Wang}, {Broderick},
  {Doeleman}, {Fish}, {Loeb}, \& {Psaltis}}]{Johannsen2016}
{Johannsen}, T., {Wang}, C., {Broderick}, A.~E., {et~al.} 2016, \prl, 117,
  091101, \dodoi{10.1103/PhysRevLett.117.091101}

\bibitem[{{Joshi}(2012)}]{2012gcss.book}
{Joshi}, P.~S. 2012, {Gravitational Collapse and Spacetime Singularities}

\bibitem[{{Kauffmann} \& {Heckman}(2009)}]{Kauffmann2009}
{Kauffmann}, G., \& {Heckman}, T.~M. 2009, \mnras, 397, 135,
  \dodoi{10.1111/j.1365-2966.2009.14960.x}

\bibitem[{{Kumar} {et~al.}(2023){Kumar}, {Bhatt}, \&
  {Bhattacharyya}}]{Kumar2023}
{Kumar}, R., {Bhatt}, N., \& {Bhattacharyya}, S. 2023, \mnras, 524, L55,
  \dodoi{10.1093/mnrasl/slad065}

\bibitem[{{Lapi} {et~al.}(2018){Lapi}, {Pantoni}, {Zanisi}, {Shi}, {Mancuso},
  {Massardi}, {Shankar}, {Bressan}, \& {Danese}}]{Lapi2018}
{Lapi}, A., {Pantoni}, L., {Zanisi}, L., {et~al.} 2018, \apj, 857, 22,
  \dodoi{10.3847/1538-4357/aab6af}

\bibitem[{{Liu} {et~al.}(2021{\natexlab{a}}){Liu}, {Zhu}, \& {Wu}}]{Liu1}
{Liu}, C., {Zhu}, T., \& {Wu}, Q. 2021{\natexlab{a}}, Chinese Physics C, 45,
  015105, \dodoi{10.1088/1674-1137/abc16c}

\bibitem[{{Liu} {et~al.}(2021{\natexlab{b}}){Liu}, {Zhu}, \& {Wu}}]{Liu2021}
---. 2021{\natexlab{b}}, Chinese Physics C, 45, 015105,
  \dodoi{10.1088/1674-1137/abc16c}

\bibitem[{{Long} {et~al.}(2022){Long}, {Yang}, {Chen}, \& {Wang}}]{Long2022}
{Long}, F., {Yang}, M., {Chen}, J., \& {Wang}, Y. 2022, International Journal
  of Modern Physics A, 37, 2250206, \dodoi{10.1142/S0217751X22502062}

\bibitem[{{Maurya} {et~al.}(2022){Maurya}, {Govender}, {Singh}, \&
  {Nag}}]{Maurya2022}
{Maurya}, S.~K., {Govender}, M., {Singh}, K.~N., \& {Nag}, R. 2022, European
  Physical Journal C, 82, 49, \dodoi{10.1140/epjc/s10052-021-09979-9}

\bibitem[{{Merloni} \& {Heinz}(2013)}]{Merloni2013}
{Merloni}, A., \& {Heinz}, S. 2013, in Planets, Stars and Stellar Systems.
  Volume 6: Extragalactic Astronomy and Cosmology, ed. T.~D. {Oswalt} \& W.~C.
  {Keel}, Vol.~6, 503, \dodoi{10.1007/978-94-007-5609-0_11}

\bibitem[{{Misner} {et~al.}(1973){Misner}, {Thorne}, \& {Wheeler}}]{Misner1973}
{Misner}, C.~W., {Thorne}, K.~S., \& {Wheeler}, J.~A. 1973, {Gravitation}

\bibitem[{{Mountrichas}(2023)}]{Mountrichas2023}
{Mountrichas}, G. 2023, \aap, 672, A98, \dodoi{10.1051/0004-6361/202345924}

\bibitem[{{Narayan} \& {McClintock}(2008)}]{Narayan2008}
{Narayan}, R., \& {McClintock}, J.~E. 2008, \nar, 51, 733,
  \dodoi{10.1016/j.newar.2008.03.002}

\bibitem[{{Narayan} \& {McClintock}(2013)}]{Narayan2013}
---. 2013, arXiv e-prints, arXiv:1312.6698, \dodoi{10.48550/arXiv.1312.6698}

\bibitem[{{Natarajan}(2004)}]{Natarajan2004}
{Natarajan}, P. 2004, in Astrophysics and Space Science Library, Vol. 308,
  Supermassive Black Holes in the Distant Universe, ed. A.~J. {Barger}, 127,
  \dodoi{10.1007/978-1-4020-2471-9_4}

\bibitem[{{Rees}(1984)}]{Rees1984}
{Rees}, M.~J. 1984, \araa, 22, 471, \dodoi{10.1146/annurev.aa.22.090184.002351}

\bibitem[{{Rezzolla} \& {Zanotti}(2013)}]{Rezzolla2013}
{Rezzolla}, L., \& {Zanotti}, O. 2013, {Relativistic Hydrodynamics}

\bibitem[{{Schutz}(2009)}]{Schutz2009}
{Schutz}, B. 2009, {A First Course in General Relativity}

\bibitem[{{Volonteri} {et~al.}(2021){Volonteri}, {Habouzit}, \&
  {Colpi}}]{Volonteri1}
{Volonteri}, M., {Habouzit}, M., \& {Colpi}, M. 2021, Nature Reviews Physics,
  3, 732, \dodoi{10.1038/s42254-021-00364-9}

\bibitem[{{Wei} \& {Liu}(2021)}]{Wei2021EPJP}
{Wei}, S.-W., \& {Liu}, Y.-X. 2021, European Physical Journal Plus, 136, 436,
  \dodoi{10.1140/epjp/s13360-021-01398-9}

\bibitem[{{Woods} {et~al.}(2019){Woods}, {Agarwal}, {Bromm}, {Bunker}, {Chen},
  {Chon}, {Ferrara}, {Glover}, {Haemmerl{\'e}}, {Haiman}, {Hartwig}, {Heger},
  {Hirano}, {Hosokawa}, {Inayoshi}, {Klessen}, {Kobayashi}, {Koliopanos},
  {Latif}, {Li}, {Mayer}, {Mezcua}, {Natarajan}, {Pacucci}, {Rees}, {Regan},
  {Sakurai}, {Salvadori}, {Schneider}, {Surace}, {Tanaka}, {Whalen}, \&
  {Yoshida}}]{Woods1}
{Woods}, T.~E., {Agarwal}, B., {Bromm}, V., {et~al.} 2019, \pasa, 36, e027,
  \dodoi{10.1017/pasa.2019.14}

\bibitem[{{Zhu} {et~al.}(2022){Zhu}, {Li}, {Li}, {Maji}, {Yajima}, {Schneider},
  \& {Hernquist}}]{Zhu1}
{Zhu}, Q., {Li}, Y., {Li}, Y., {et~al.} 2022, \mnras, 514, 5583,
  \dodoi{10.1093/mnras/stac1556}

\end{thebibliography}

\end{document}